\documentclass{emulateapj}
\usepackage{graphicx}
\usepackage{psfig}
\usepackage{natbib}

\newcommand{\gsim}{\mbox{\hspace{.2em}\raisebox{.5ex}{$>$}\hspace{-.8em}\raisebox{-.5ex}{$\sim$}\hspace{.2em}}}
\newcommand{\lsim}{\mbox{\hspace{.2em}\raisebox{.5ex}{$<$}\hspace{-.8em}\raisebox{-.5ex}{$\sim$}\hspace{.2em}}}
\newcommand{\RA}[3]{{#1}^{{\rm h}}{#2}^{{\rm m}}{#3}^{{\rm s}}}
\newcommand{\Dec}[3]{{#1}^{\circ}{#2}'{#3}''}
\newcommand{\E}[1]{\times 10^{#1}}
\newcommand{\twCO}{$^{12}$CO}  \newcommand{\thCO}{$^{13}$CO}
\newcommand{\HII}{\mbox{H\,\textsc{ii}}}

      \newcommand{\ps}{\,{\rm s}^{-1}}
    \newcommand{\Msun}{M_{\odot}}   
    \newcommand{\km}{\,{\rm km}}

\begin{document}

\title{
The Distant Outer Gas Arm Between $35^{\circ}$ and $45^{\circ}$ 
}

\shorttitle{The Distant Outer Gas Arm}

\author{
Yang Su\altaffilmark{1}, Yan Sun\altaffilmark{1}, 
Chong Li\altaffilmark{1,2},
Shaobo Zhang\altaffilmark{1}, Xin Zhou\altaffilmark{1},
Min Fang\altaffilmark{1}, Ji Yang\altaffilmark{1} 
and Xuepeng Chen\altaffilmark{1}
       }

\affil{
$^1$ Purple Mountain Observatory and Key Laboratory of Radio Astronomy, 
Chinese Academy of Sciences, Nanjing 210008, China \\
$^2$ Graduate University of the Chinese Academy of Sciences, 19A Yuquan Road, 
Shijingshan District, Beijing 100049, China \\
      }

\begin{abstract}
The Galactic plane has been mapped from $l=$34\fdg75 to 45\fdg25
and $b=-$5\fdg25 to 5\fdg25 in the CO ($J$=1--0) emission
with the 13.7 m telescope of the Purple Mountain Observatory.
The unbiased survey covers a large area of 110 square degrees
sampled every 30$''$ with a velocity resolution of $\sim0.2\km\ps$.
In this paper, we present the result of 
an unbiased CO survey of this longitude and latitude range 
in the velocity range from $-60$ to $-10\km\ps$. 
Over 500
molecular clouds (MCs) are picked out from the \twCO\ ($J$=1--0) emission, 
and 131 of these MCs are associated with \thCO\ emission.
The distant MCs, which lie beyond the solar circle and are mostly 
concentrated in the Galactic plane, 
trace the large-scale molecular gas structure over
10 degrees of Galactic azimuth.
We find that the distribution of the distant MCs 
can be well fitted by a Gaussian function
with a full width at half maximum (FWHM) of 0\fdg7 with the
Galactic latitude.
We suggest that the CO emission of the segment is from the Outer Arm.
The physical mid-plane traced by the Outer Arm 
seems to be slightly displaced from the IAU-defined plane on a large scale, 
which could be explained by the warped plane at large Galactocentric 
distances of $\gsim$10 kpc and the 
apparent tilted mid-plane to the projected IAU-defined plane 
caused by the Sun's $z$-height above the disk for distances
near and within the Solar circle. After removing the effect of the warp 
and tilted structure, the scale height of the 
MCs in the Outer Arm is about 0\fdg6 or 160 pc at a heliocentric distance 
of 15 kpc. If the inner plane of our Galaxy is flat,
we can derive an upper limit of the Sun's offset of 
$\sim$17.1 pc above the physical mid-plane of the Milky Way. 
We also discuss the correlations between the physical 
parameters of the distant MCs, which is quite consistent with 
the result of other studies of this parameter.
\end{abstract}


\section{INTRODUCTION}
Molecular clouds (MCs), as observed in CO surveys of the 
Galactic plane, play an important role in studying star formation
and the structure of the Galaxy \citep[e.g.,][]{1985ApJ...295..422C,1987ApJ...322..706D,
2001ApJ...547..792D,1998ApJ...502..265H,2006ApJS..163..145J,2013PASA...30...44B}.
Several works were concentrated on the large-scale structure traced by 
MCs in the first Galactic quadrant 
\citep[e.g.,][]{1985ApJ...297..751D,1986ApJS...60..297C,
1988ApJ...327..139C,1986ApJ...305..892D,1987ApJS...63..821S,1988A&A...195...93J,1989ApJ...339..919S,
2010ApJ...723..492R}.
However, the CO emission lying beyond the solar circle in the first quadrant 
of the Galaxy is less studied \citep[e.g., Section 3.1.5 and 
Figure 3 in][]{2001ApJ...547..792D}.  
The only cases are the study of the MCs 
in the Outer Arm of the Galaxy for the region of $l>55^{\circ}$ 
in the first quadrant \citep[using the NRAO 11 m telescope,]
[]{1981ApJ...249L..15K,1988ApJ...330..399M} 
and for the region between $l=65^{\circ}$ and $71^{\circ}$ 
\citep[using the CfA 1.2 m and the Bell 7 m telescopes,][]{1990ApJ...357L..29D}.

The Milky Way Imaging Scroll Painting (MWISP) project 
\footnote{http://www.radioast.nsdc.cn/mwisp.php} is 
a high resolution ($50''$) \twCO\ ($J$=1--0), \thCO\ ($J$=1--0), and C$^{18}$O ($J$=1--0) 
survey of the northern Galactic plane, performed with the 
Purple Mountain Observatory Delingha 13.7 m telescope.
The survey started in 2011, and will cover Galactic longitudes from
$l=-10^{\circ}$ to $250^{\circ}$ and latitudes from 
$b=-5^{\circ}$ to $5^{\circ}$ over a period of $\sim$10 years.
The Galactic plane will be covered by full-sampling observations 
with the spectral line On-The-Fly (OTF) mode. The survey has equal
sensitivity over all regions of the sky mapped and thus it is unbiased.
One of the goals of this project is to study the physical properties of MCs
along the northern Galactic plane. 
As of this writing the survey has completed about half of its planned 
area of coverage. 

Benefiting from the large unbiased 
survey of the MWISP, we can systematically investigate the 
characteristics of the structure of the Galaxy using the molecular gas. 
Recently, a CO spiral arm lying beyond the Outer Arm in the first
Galactic quadrant was identified by \cite{2011ApJ...734L..24D}
using the CfA 1.2 m telescope.
Based on our new CO data of the MWISP, the extension of the above arm, 
which is probably the Scutum-Centaurus Arm into the outer second quadrant,
has been revealed very recently by \cite{2015ApJ...798L..27S}.

In this paper, we investigate the Outer Arm 
\citep[the Norma-Cygnus Arm, e.g., see the recent review in][]{2014ApJS..215....1V}
in the first quadrant of the Galaxy according to the distribution 
of the CO gas between $l=35^{\circ}$ to $45^{\circ}$.
The distant MCs beyond the solar circle traced by CO emission have a negative velocity
in the direction that we are interested in. The velocity range
of the CO gas within the arm is about $-60$ to $-10\km\ps$ 
in the direction (Section 3.3). 
The Outer Arm lies beyond the solar circle in the Galactic longitude
range of $l\gsim30^{\circ}$ in the first quadrant of the Milky Way, 
thus the molecular gas traced by CO emission does not suffer the 
kinematic distance ambiguity encountered in the region within the Solar circle.
Generally speaking, in this quadrant, the more negative 
the value of the velocity is, the larger the distance of the MC.
Moreover, the large sample of the distant MCs provides a good opportunity 
to study the global cloud parameters (Section 3.6) since the data 
have a high signal-to-noise ratio and there is little overlap 
between the clouds (Section 3.1).

We present observations and data reductions of the survey in Section 2.
In Section 3.1 and 3.2, we describe the MCs identification and
the distribution of the distant MCs, respectively.
We find that the physical mid-plane traced by the distant CO
gas is slightly displaced from the IAU-defined plane of $b = 0^{\circ}$
(Section 3.3). We also discuss the relationship between 
the two planes in Section 3.4. In Section 3.5, we derive the Sun's 
offset above the physical mid-plane because of the apparent displacement 
between the two planes for regions near and within the solar circle.
The statistical properties of the distant MCs are 
discussed in Section 3.6.
Finally, the summary is given in Section 4.

\section{CO OBSERVATIONS AND DATA REDUCTIONS}
We mapped $\sim$110 square degrees (439 cells) in the 
region between Galactic longitudes 34\fdg75--45\fdg25 and 
latitudes $-$5\fdg25 to 5\fdg25 during 2011 November to
2015 March using the 13.7 m millimeter-wavelength telescope located 
at Delingha in China.
The nine-beam Superconducting Spectroscopic Array Receiver (SSAR) 
system was used at the front end and each Fast Fourier transform 
spectrometer (FFTS) with a bandwidth of 1 GHz provides 16,384 channels 
and a spectral resolution of 61 kHz \citep[see the details in][]{Shan}.
The molecular lines of \twCO\ ($J$=1--0), \thCO\ ($J$=1--0) and
C$^{18}$O ($J$=1--0) were observed simultaneously with the OTF method. 
Each cell with dimension 30$'\times30'$
was scanned at least in two orthogonal directions, along the Galactic 
longitude and the Galactic latitude, in order to reduce the fluctuation 
of noise. The half-power beamwidth (HPBW) of the telescope
was about $50''$ at a local oscillating frequency of 112.6 GHz 
and the pointing accuracy was better than $5''$ in all observing epochs.

All cells were reduced using the GILDAS/CLASS
package.\footnote{http://www.iram.fr/IRAMFR/GILDAS}
After removing the bad channels in the spectra and correcting the
first order (linear) baseline fitting, the antenna temperature
($T_{\rm A}^{\star}$) was converted to the main beam temperature
($T_{\rm mb}$) with the relation 
$T_{\rm mb}=T_{\rm A}^{\star}/(f_{\rm b}\times\eta_{\rm mb})$.
In the above conversion, the beam-filling factor of $f_{\rm b}$ is 
assumed to be one and the beam efficiency $\eta_{\rm mb}$ is from the 
status report\footnote{http://www.radioast.nsdc.cn/mwisp.php} 
of the 13.7 m telescope of the Purple Mountain Observatory.
The typical sensitivity (rms) is about 0.5 K for \twCO\ ($J$=1--0) 
at the channel width of 0.16$\km\ps$ and 0.3 K for \thCO\ ($J$=1--0) and 
C$^{18}$O ($J$=1--0) at 0.17$\km\ps$ 
with a spatial resolution of $<1'$.

Finally, all spectra of CO in each cell were converted to the 
three-dimensional (3D) cube mosaic data together with a grid 
spacing of $30''$ and the channel width of 0.16$\km\ps$ for 
\twCO\ ($J$=1--0) and 0.17$\km\ps$ for \thCO\ ($J$=1--0) 
and C$^{18}$O ($J$=1--0) for subsequent analysis.

\section{RESULTS AND DISCUSSION}
\subsection{MCs Identification}
We carefully checked the 3D cube data channel by channel 
before investigating
the characters of the MCs that we are interested in. We find
that the distant MCs in the velocity interval of 
$-60$ to $-10\km\ps$ 
are isolated in the $l$--$b$--$v$ space. 
That is, most of them show the distinct structure in the space of 
$l$--$b$ and there is, therefore, little overlapping in the 
velocity axis over the large survey's area. Moreover, it is 
not surprising that their \twCO\ ($J$=1--0) emission
is stronger than that of the \thCO\ ($J$=1--0).
No $>3\times$rms C$^{18}$O ($J$=1--0) emission 
($T_{\rm peak18}$) is detected in the whole region
in the velocity range.
Only insignificant C$^{18}$O emission ($T_{\rm peak18}\sim2\times$rms) 
can be found toward the MCs' region with some detectable 
\thCO\ emission (e.g., $T_{\rm peak13}>5\times$rms). We do not discuss these
weak C$^{18}$O emission features in this paper.

According to the discrete feature of the distant MCs,
we used the FINDCLUMPS tool in the CUPID package
(part of the STARLINK package) to identify MCs in the 
\twCO\ and \thCO\ FITS cube. The CLUMPFIND algorithm
\citep{1994ApJ...428..693W} is applied in the process of
identification. We set the parameter Tlow=3$\times$rms 
for \twCO\ and Tlow=2.7$\times$rms for \thCO,
which determines the lowest level to contour an MC, in
order to obtain as much of the emission as possible
and to avoid the contamination of the noise based on
our large survey data. The parameter of DeltaT=9$\times$rms, 
which shows the contour increment, is large since we do not 
care about the properties of the structure of the bright 
parts embedded in the extended MCs. 
Other input parameters of the CLUMPFIND algorithm are 
typical to identify those distant MCs for the 3D cube data.
Considering the irregular shape of the MCs, we use
a polygon to describe the MC's boundary in the $l$--$b$ space.
Some objects are rejected if their sizes 
are less than the criteria: FwhmBeam=1.5 pixel, 
VeloRes=2 channel, and MinPix=16. 

It is worthwhile to note that sometimes an individual MC is 
decomposed to several MCs because of the small velocity 
separation of $\sim1-2\km\ps$ within roughly the same $l$--$b$ 
region. However, the above case is a rare population compared 
to that of all detected MCs (Figures 1 and 2). 
On the other hand, some point-like sources ($\lsim$ 3--4 pixels) 
cannot be picked out because of their smaller sizes. Some faint 
diffuse structures where \twCO\ ($J$=1--0) emission is near 
the noise level of the survey cannot be identified
due to their poor signal-to-noise ratio. Further observations 
are needed to confirm these MCs with small size or faint emission.

Finally, we identified 575 and 131 MCs from the \twCO\ 
and \thCO\ emission based on the CLUMPFIND algorithm, respectively.
The parameters of each MC, 
such as position, LSR velocity, one-dimensional velocity dispersion, 
the peak value, the size, and the luminosity, were obtained 
directly from the automated detection routine. The parameters 
of the resolved MCs are summarized in Tables 1 and 2: 
(1) the ID of the identified MCs, arranged from the low
Galactic longitude; 
(2) and (3) the MCs' Galactic coordinates ($l$ and $b$);
(4) the MCs' LSR velocity ($V_{\rm LSR}$);
(5) the MCs' full width at half maximum (FWHM, $\Delta V_{\rm FWHM}$), 
defined as 2.355 times of the velocity dispersion 
$\sigma_{v}$ for a Gaussian line;
(6) the MCs' peak emission ($T_{\rm peak}$);
(7) the MCs' area;
(8) the MCs' integrated intensity ($W_{\rm CO}$) within a defined 
area;
(9) the MCs' total luminosity ($L_{\rm CO}=W_{\rm CO}\times \rm {Area}$).
Finally, we use MWISP Glll.lll$\pm$bb.bbb to name an
MC detected from the survey data.

We made the integrated intensity map of the \twCO\ 
($J$=1--0) emission in the velocity interval of 
$-1.6\times l$+$13.2\km\ps$ to $-10\km\ps$ (Section 3.3)
in Figure 1.
After overlapping the 575 resolved MCs on the map, we find that 
the distribution of the CO gas is well traced by the result from
the automated cloud-finding routine. It shows that the CLUMPFIND 
algorithm is thus reliable to pick out the distant MCs.
In addition, the sample of the MCs can be used to 
investigate the large structures of the spiral arm (Section 3.3) 
and the properties of the distant MCs (Section 3.6).

Figure 2 displays the intensity-weighted
velocity (the first moment) map of the \twCO\ ($J$=1--0) emission.
The distribution of the 131 \thCO\ MCs is also consistent with 
that of the bright \twCO\ ($J$=1--0) emission.
The blue stripes at $l=$41\fdg5 and $b=-$1\fdg5 are from the 
bad channels,
which show regular oscillation along the velocity axis
with intensity of $-2$ to 3 K. The abnormal feature can be
easily distinguished from the CO emission of MCs.

We also present the typical spectra of four resolved MCs
(\twCO\ with the black line and \thCO $\times2$ with blue) in Figure 3.
All spectra in Figure 3 show good baselines.
The CO emission of the MCs can be easily discerned from the
spectra and the peak of \thCO\ ($J$=1--0) emission is roughly 
corresponding to that of the \twCO\ ($J$=1--0) line (Figure 3).

\subsection{Spatial Distribution of the Distant Molecular Gas}
We show the 575 MCs in the Galactic coordinate system in Figure 4, 
in which the filled circle indicates the 
position and the size (scaled with CO luminosity) of MCs and the color 
indicates the LSR velocity of MCs.
All identified MCs are in the region of the Galactic latitude
$b=-$1\fdg8 to $b=$2\fdg2, which indicates that the molecular gas is
roughly concentrated in and around the Galactic plane
(mainly in the region of $b\sim -$0\fdg5 to 1\fdg5).
The map also shows that the distribution of the MCs is not 
uniform with Galactic longitude. 

Figure 5 is the normalized histogram of the MCs with the Galactic 
longitude (left panel) and the Galactic latitude (right panel). 
According to the left panel of Figure 5, we find that the spatial 
distribution 
of the MCs is not uniform again. Several distribution peaks of the
MCs are clearly discerned at $l=$35\fdg25, 39$^{\circ}$, 42$^{\circ}$, 
and 44$^{\circ}$, respectively. The concentrated groups of these MCs 
can also be seen in the same region of Figure 4.
The map in the right panel of Figure 5 shows that the MCs
are mainly located within a limited range of the Galactic latitude,
which indicates that the molecular gas in the velocity interval 
is actually from the Galactic plane with a distribution peak
of $b=$0\fdg42. 

We also made the position$-$velocity (PV) map of the MCs along
the Galactic longitude (Figure 6), 
in which the distribution of the MCs is similar 
to that in Figure 5. 
The velocity gradient of the MCs can be seen 
on a large scale (see green dashed line in Figure 6),
which shows the velocity trend of the Outer Arm 
along the Galactic longitude (Section 3.3).
The distribution of the MCs in the longitude$-$velocity 
diagram seems to display some asymmetrical ripples,
which is a feature that is probably from the substructures
of the Outer Arm in this direction. 

We note that some MCs are concentrated within groups.
These MC groups seem to display partial shell structures,
which probably relate to the star forming activity within
the Outer Arm. 
We searched for the MCs associated with the massive star
formation regions based on our unbiased CO survey. 
The information of the associations is listed in Table 3.
We find that the distribution of the massive star-forming 
regions in the Outer Arm, which are mainly located at 
$l = 39^{\circ}$, $42^{\circ}$, and $44^{\circ}$, is associated 
with that of the CO peaks in the $l$--$b$ space
(see the 21 \HII\ regions marked with black circles and 
five 6.7 GHz masers marked with black triangles in Figure 2).
This result shows that star formation activity is very common
in the distant MCs of the Outer Arm.
We will study the interesting structures and the massive 
star-forming activities in the future.

\subsection{Outer Gas Arm Traced by CO emission} 
In the direction of $l = 35^{\circ}$ to $45^{\circ}$, we note that 
the Outer Arm LSR velocity locates between the Perseus Arm
and the Scutum-Centaurus Arm. 
That is, the Perseus Arm, the Outer Arm, and the Scutum-Centaurus 
Arm are encountered in this direction, each with different 
and decreased LSR velocity 
\citep[see Figures 9 and 10 in][]{2016ApJ...823...77R}.
Part of the molecular gas' emission in the velocity interval of 
0 to $-10\km\ps$ is from the Perseus Arm. 
The molecular gas of the Local Arm
\citep{2013ApJ...769...15X} probably also contributes to some emission
in the velocity range of 0 to $-10\km\ps$. And
the molecular gas with the LSR velocity $< -1.6\times l$+$13.2\km\ps$ 
is probably from a segment of a spiral arm at
Galactocentric radii of R$_{\rm GC} \sim$13--14 kpc 
\citep[the value of -1.6 $\km\ps$Degree$^{-1}$ is from][]{2011ApJ...734L..24D}.
In order to isolate the emission from the three spiral arms, 
we use the molecular gas in the velocity cutoff of
$-1.6\times l$+$13.2\km\ps$ to $-10\km\ps$
to trace the Outer Arm (see the yellow dashed lines in Figure 6). 
Actually, the large-scale segment of the Outer Arm can be  
discerned from the longitude$-$velocity diagram of CO emission, in which 
the CO emission of the Outer Arm's MCs is mainly in the 
above velocity range and their distribution can be 
described as $V_{\rm LSR}=-2.78\times l+75.9\km\ps$ 
(see the green dashed line in Figure 6) based on
\thCO\ emission (excluding MCs near $l\sim44^{\circ}$ and
$V_{\rm LSR}\sim-20 \km\ps$).

We use a Gaussian function, $f(b)\sim e^{-\frac{(b-b_{0})^{2}}{2\sigma ^{2}}}$,
to fit the distribution of the MCs with the Galactic latitude
(the right panel in Figure 5).
The dashed blue line shows the best fit: $b_{0}$=0\fdg42 and
$\sigma$=0\fdg29. It indicates that the distant MCs in the velocity 
interval are mainly distributed within
a narrow range of the Galactic latitudes of 
FWHM=2.355$\sigma \sim$0\fdg7.
We thus suggest that the large amount of the CO emission actually 
traces the Outer Arm of the Milky Way.

The best Gaussian fit also shows that the distribution peak of 
the MCs in the segment is not around $b = 0^{\circ}$ but 
$b$=0\fdg42, which probably indicates that 
the Outer Arm traced by the molecular gas is slightly displaced 
with respect to the IAU-defined plane (see the details in Section 3.4). 
We also use a linear relation, $b$=0\fdg0377$\times l -$1\fdg0893 (the 
thick blue line in Figure 4), to describe the distribution of the MCs 
in the $l$--$b$ space. 
The fitted blue line represents the physical mid-plane with respect to
the IAU-defined plane of $b = 0^{\circ}$.
It indicates that the Outer Arm traced by the CO gas passes through 
the IAU-defined plane of $b = 0^{\circ}$ at $l = $28\fdg9.
The molecular gas of the Outer Arm in the range of $l\lsim$28\fdg9 is 
mostly within the solar circle and the distribution of 
the MCs with latitude is mainly $b< 0^{\circ}$ 
\citep[see the Norma-Cygnus arm in Figure 3 of][]{2014ApJS..215....1V}.
The longitude$-$velocity relation from Figure 6 also indicates 
that the LSR velocity 
of the Outer Arm is roughly $\gsim0 \km\ps$ at $l\lsim$28\fdg9.
The feature is probably related to the 
Sun's offset from the physical mid-plane (see Sections 3.4 and 3.5).

Using the 21 cm \mbox{H\,\textsc{i}}
data from the VLA Galactic plane survey
\citep[VGPS;][]{2006AJ....132.1158S},
\cite{2011ApJ...734L..24D} found that the Outer Arm
traced by the \mbox{H\,\textsc{i}} gas is mainly 
below the plane of $b = 0^{\circ}$ in the region
of $l = 25^{\circ}-30^{\circ}$ (see Figure 2 in their paper). 
In other words, at $l < 30^{\circ}$ the Outer Arm seems to be
below the IAU-defined plane in the first Galactic quadrant.
According to our linear fit of the CO emission, in the region of 
$l = 25^{\circ}-30^{\circ}$ the Galactic latitude distribution of 
the Outer Arm should peak at $b \sim-$0\fdg1, which is 
roughly consistent with the result of \mbox{H\,\textsc{i}} gas 
\citep[see Figure 2 in][]{2011ApJ...734L..24D}.

Based on the Galactic parameters of Model A5 of \cite{2014ApJ...783..130R}, 
we suggest that the heliocentric distance of the 10$^{\circ}$ segment of the
Outer Arm is about 15.1--15.6 kpc, which is a value that is 
quite consistent with the result of other 
models \citep[e.g.,][]{2007ApJ...671..427M,2010ASPC..438...16F}.
The Galactocentric radii of the arm, R$_{\rm GC}$, in the direction is 
about 9.5--11.4 kpc accordingly.
After subtracting the effect of the tilted structure with the 
Galactic longitude (the thick blue line in Figure 4), we derived  
that the scale height of the MCs in the Outer Arm is about 
0\fdg6 (or $\sim$160 pc at a heliocentric distance of 15 kpc).
It is roughly consistent with the FWHM of the molecular disk of the 
Milky Way in such Galactocentric radii 
\citep[see Table 1 and Figure 3 in][]{2006PASJ...58..847N}.

The total \twCO\ luminosity of the segment of the spiral arm
is about 2.1$\times10^5$ K$\km\ps$pc$^2$ at a heliocentric distance of 15 kpc.
Therefore, the total molecular gas within the segment is
about 9$\times10^{5}\Msun$ by adopting the mean CO-to-H$_2$ mass 
conversion factor $X_{\rm CO}$=$2\E{20}$~cm$^{-2}$K$^{-1}$km$^{-1}$s 
\citep{2013ARA&A..51..207B} and a mean molecular weight per
H$_2$ molecule of 2.76.
We mention that the total mass of the segment estimated above 
is probably the lower limit because some weak \twCO\
emission is not accounted for in Table 1 (Section 3.1)
and the value of $X_{\rm CO}$ is probably underestimated for the outer 
Milky Way \citep{2013ARA&A..51..207B}.

Based on Table 1, the \twCO\ luminosity of the segment
of the spiral arm is about 2.1$\times10^5$ K$\km\ps$pc$^2$ at a
heliocentric distance of 15 kpc.
The molecular gas within the segment is
about 0.9$\times10^{6}\Msun$ by adopting the mean CO-to-H$_2$ mass
conversion factor $X_{\rm CO}$=$2\E{20}$~cm$^{-2}$K$^{-1}$km$^{-1}$s
\citep{2013ARA&A..51..207B} and a mean molecular weight per
H$_2$ molecule of 2.76.
For comparison, the total \twCO\ luminosity of the segment
of $\sim2.6\times10^4$ K$\km\ps$arcmin$^2$ can be obtained by
integrating velocity channels from $-1.6\times l$+$13.2\km\ps$
to $-10\km\ps$ for the covered map. Therefore, our estimate for 
the total molecular mass of the segment of the Outer Arm
is 2.2$\times10^{6}\Msun$. The result indicates that
over half of the total mass appears to be missed due to the relatively
high criterion of Tlow=3$\times$rms in the CLUMPFIND method.
The difference between the above estimations probably indicates
that about 60\% of the molecular gas probably resides within small, cold,
and$\//$or faint, diffuse clouds that cannot be included in our catalog
\citep[see also, e.g., Section 5.1.3 of][]{2015ARA&A..53..583H}.
We mention that the total mass of the segment estimated above
is probably the lower limit because some weak \twCO\
emission is not accounted for and the value of $X_{\rm CO}$ is
probably underestimated for the outer
Milky Way \citep{2013ARA&A..51..207B}.
Assuming a constant value of $X_{\rm CO}$=
$2\E{20}$~cm$^{-2}$K$^{-1}$km$^{-1}$s and a heliocentric distance of 15 kpc, 
the mass surface density of
the distant MCs is about $22\Msun$~pc$^{-2}$, which is roughly consistent with
the character of MCs in the outer Galaxy 
\citep[see, e.g., Figure 8 in][]{2015ARA&A..53..583H}.

\subsection{Physical Mid-plane Traced by the Outer Arm} 
In this section, we investigate the relationship between
the physical mid-plane and the IAU-defined plane 
of $b = 0^{\circ}$ in the view
of the molecular gas on a large scale based on the new CO survey.

Generally, warps are common in spiral disks of galaxies
\citep[e.g.][]{1998A&A...337....9R,2001A&A...373..402S,2002A&A...382..513R}. For example,
NGC 4013 shows a prodigious warp in its outer parts,
in which the gas layer curves away from the plane of the inner
disk \citep{1996A&A...306..345B}. In the Milky Way, the H\textsc{i}
layer was also found to be warped from the Solar radius outwards
\citep[e.g.,][]{1957AJ.....62...90B,1958MNRAS.118..379O,1982ApJ...263..116H}.
Furthermore, the stellar warp of the Milky Way is similar to the gaseous
warp, but smaller in amplitude
\citep[e.g.,][]{1994ApJ...429L..69F,1998ApJ...492..495F,2001ApJ...556..181D}.

On the other hand, the inner disk of the Milky Way is probably
observed to be
tilting because of the offset of the Sun \citep{1995MNRAS.273..206H}. 
That is, the physical mid-plane projected on the sky displays an apparent 
change in latitude with respect to longitude across some regions of the sky.
Previously, astronomers often used the star-counts 
method to derive the position of the Sun
\citep[e.g.,][]{1995MNRAS.273..206H,1995AJ....110.2183H,2001ApJ...553..184C,
2001AJ....121.2737M,2007MNRAS.378..768J,2009MNRAS.398..263M}.
These studies found
that the Sun's vertical displacement from the physical mid-plane is 
$\sim$10--30 pc \citep[refer to Table 1 in][]{2006JRASC.100..146R}.
By studying the infrared dark cloud (IRDC) ``Nessie,''
\cite{2014ApJ...797...53G} recently suggested that the very long 
and dense filamentary IRDC represents a spine-like bone of the
major spiral arm, the Scutum-Centaurus Arm, in the fourth quadrant
of the Milky Way. They also suggested that the latitude 
of the true Galactic mid-plane traced by the spiral arm is 
$b\sim -$0\fdg4 (but not $b=0^{\circ}$)
because of the Sun's height off of the physical mid-plane 
of the Milky Way 
\citep[see Figure 2 and Section 3.1 in][]{2014ApJ...797...53G}.

Here we discuss the Galactic structure traced by CO emission of the
Outer Arm. 
Both of the effects of the warp and the tilted plane (because of the Sun's 
offset from the physical mid-plane) were considered.
We explored three typical models for comparison with our observations:
the gaseous warp model from the Galactic H\textsc{i} data
\citep[$m=1$ mode for a cutoff at R$_{\rm GC}$=10 kpc,][]{2006ApJ...643..881L};
the stellar warp model from 2MASS data
\citep[a $\gamma_{\rm warp}$ of 0.09 and a cutoff 
at R$_{\rm GC}$=8.4 kpc,][]{2009A&A...495..819R};
and the tilted-plane model for the inner Galaxy
\citep[R$_{\rm GC}<7$ kpc, and see Equation 5 in][]{1995MNRAS.273..206H}.
The $\gamma_{\rm warp}$ in the stellar warp model is defined as 
the ratio between the displacement (z$_{\rm mid-plane}$)
of the physical mid-plane from the IAU-defined plane of $b = 0^{\circ}$
and the Galactocentric radii of R$_{\rm GC}$
\citep[refer to Section 2.3 and Figure 9 in][]{2009A&A...495..819R}.
For two warp models, we assume that the Sun roughly lies on the line of 
nodes of the warp 
\citep[i.e., $\phi_{\rm warp}=0^{\circ}$,][]{1988gera.book..295B,1998gaas.book.....B}.
Based on the best fit of the longitude$-$velocity diagram 
($V_{\rm LSR}=-2.78\times l+75.9\km\ps$; Figure 6), we can 
kinematically calculate the Galactocentric radii and the height 
of the physical mid-plane traced by the Outer Arm using the A5 
rotation curve model of \cite{2014ApJ...783..130R}.
Thus, when the R$_{\rm GC}$--z$_{\rm mid-plane}$ relation is obtained, 
we can directly compare the slope and the amplitude of the different 
models (the dotted lines) with those of the observations (the blue line) 
in the $l-b$ map (Figure 4).
Here, the slope is defined as the ratio between 
z$_{\rm mid-plane}$ and $\Delta l$.

We find that the slopes of the two 
warp models are larger than that of the observations.
The amplitude of the stellar warp model is larger than the 
observed displacement across the present longitude interval. 
Furthermore, the amplitude of the gaseous warp model is also larger
than the observations in the region of $l\gsim$40\fdg7 or R$_{\rm GC}\gsim10.6$ kpc.
It indicates that the effect of the warp 
traced by the CO gas of the Outer Arm is $not$ so large in 
this segment with R$_{\rm GC}\sim9.5-11.4$ kpc.
This result is roughly consistent with the interpretation that
the amplitude of the physical mid-plane displacement is small inside a radius of 10 kpc, 
and it steeply increases beyond the radius \citep{2006PASJ...58..847N}.
Moreover, \cite{2002A&A...394..883L} also suggested that the effect
of the stellar warp is large at the place of the larger R$_{\rm GC}$,
but is small in the vicinity of the solar circle
(R$_{\rm GC}\sim$ 8--10 kpc, see Figure 18 in their paper).
Our CO survey shows that the slope of the observations is intermediate
between the warp models and the tilted flat plane model 
across the region.
Therefore, the warp probably affects the observed slope of the MCs 
in such a region, but not to the extent predicted by the current models.

On the contrary, the slope of the tilted-plane model
seems very close to that of the observations 
in the region of $l\lsim$37\fdg5 (or R$_{\rm GC}\lsim10$ kpc). 
The displacement between the physical mid-plane and the $b=0^{\circ}$ plane
is also comparable to that of the tilted-plane model in the above region, 
in which the offset between the two planes is 
about 0\fdg2--0\fdg3 or 50--80 pc at a heliocentric distance of 15 kpc (Figure 4). 
Furthermore, the amplitude of the tilted-plane model will be closing to 
the displacement between the two planes with decreased R$_{\rm GC}$ 
(or $l$ for the mid-plane traced by the Outer Arm).
Actually, the inner H\textsc{i} disk was found to be tilted against 
the $b = 0^{\circ}$ plane \citep{2003PASJ...55..191N}. Also,
the amplitude of the displacement traced by our CO gas seems comparable to
that of \cite{2006PASJ...58..847N} (see Figure 12 in 
their paper) in the region with R$_{\rm GC}\sim$9--11 kpc.

Based on the above analysis, we suggest that the Galactic warp plays a role 
in the region with large Galactocentric radii (e.g., R$_{\rm GC}\gsim10$ kpc)
while it could be neglected in regions near and within the solar circle.
By contrast, the tilted-plane model
suggests that the Sun's offset from the 
physical mid-plane seems to become the dominant for regions 
near and within the solar circle.
Roughly speaking, to first order approximation, the observed tilt of 
the MCs is due to the Sun's offset from the physical mid-plane
for regions with R$_{\rm GC}\lsim10$ kpc.

\subsection{Position of the Sun}


In Section 3.4, we show that the Galactic warp is not strong
and the apparent displacement between the two planes is probably dominated by
the tilted-plane effect in the region of R$_{\rm GC}\lsim10$ kpc.
We thus do not take into account the effect of the gas warp in the region 
near and within the solar circle.
Indeed, the Galactic warp is very important for the distant gas arm with large 
R$_{\rm GC}$, \citep[e.g., see that the structure of the 
extension of the Scutum-Centaurus Arm with R$_{\rm GC} >14$ kpc 
exhibits warps along the Galactic longitude, Figure 2b in]
[]{2015ApJ...798L..27S}. 
The warp of the Outer arm is also obvious between 
$l=100^{\circ}$ and $150^{\circ}$ \citep[Figure 7 in][]{2016ApJS..224....7D},
in which the Galactocentric radii ($\gsim 13$ kpc) of the
Outer Arm is larger than that of the MCs discussed by us here.

According to Section 3.4, to first order, the displacement 
between the physical mid-plane and the $b = 0^{\circ}$ plane is 
dominated by the 
tilted effect because of the Sun's height off of the physical mid-plane 
for regions of R$_{\rm GC}\lsim10$ kpc.
If the tilt of the Galactic plane is the cause of the 
observations here, then we can derive the offset of the Sun
by extrapolating the fitted Outer Arm into the lower longitude 
(e.g., $l\sim25^{\circ}-35^{\circ}$).
In Figure 7, we construct a schematic diagram to show the 
presumable relationship between the physical mid-plane
and the IAU-defined plane within the solar circle.

In Figure 7, $\alpha$=$-$0\fdg046 is the Galactic latitude
of Sgr A$^{\star}$ \citep[$(\RA{17}{45}{40}.0409,\Dec{-29}{00}{28}.118)$ in
J2000 for Sgr A$^{\star}$,][]{2004ApJ...616..872R} and
$\beta$=28\fdg9 is the Galactic longitude of the Outer Arm 
at $b=0^{\circ}$ (Section 3.3).
The tilted angle, $\theta$, is to be determined from the 
geometric relation shown in the schematic diagram. That is, the
Outer Arm passes through the IAU-defined plane
at the direction of $l = \beta$ and $b = 0^{\circ}$.
The value for the fitted parameter of 
$\beta =l = $28\fdg9 at $b=0^{\circ}$ 
can be derived from the linear fit (the blue solid line in Figure 4).
Assuming the distance of $d_{\rm Sgr A^{\star}}=$8.34 kpc
\citep{2014ApJ...783..130R}, the value of $d_{\rm Crossing}$
is about 13.69 kpc (see the lower left corner of Figure 7).
Accordingly, the tilted angle $\theta$ and the Sun's offset
above the physical mid-plane $Z_{\rm Sun}$ is about 
0\fdg072 and 17.1 pc, respectively.

In our geometrical model (Figure 7), the values of $\theta$ 
and $Z_{\rm Sun}$ that we are interested in are dependent on 
$\beta$ and the distance to the Galactic center, 
$d_{\rm Sgr A^{\star}}$. We note that the tilted angle 
($\theta$) may slightly vary from $\sim$0.067 to $\sim$0.078 when 
$\beta$ is from 27$^{\circ}$
to 31$^{\circ}$. The offset of the Sun ($Z_{\rm Sun}$) 
may also vary from 15.8 pc to 18.4 pc when 
$d_{\rm Sgr A^{\star}}$ is changed from 8.0 kpc to 8.5 kpc and
the $\beta$ range of 27$^{\circ}$--31$^{\circ}$.
Comparing the value of $Z_{\rm Sun}$ with other investigations
\citep[e.g., Table 1 in][]{2006JRASC.100..146R}, our 
estimation of $Z_{\rm Sun}$=15.8--18.4 pc 
based on the distant large-scale molecular gas
agrees well with those from the method of star counts.
Moreover, the above estimation from the \twCO\ 
emission (Table 1) is similar to that from the \thCO\ data 
(Table 2; $b$=0\fdg0335$\times l -$0\fdg9640 and $l = \beta$=28\fdg8
at $b = 0^{\circ}$).
Finally, since the warp was neglected in regions
near and within the Solar circle (R$_{\rm GC}\lsim10$ kpc), the Sun's 
$z$-height based on the tilted appearance of the distant MCs 
represents an upper limit.

It should be noted that our result is slightly larger
than that of \cite{1993A&A...275...67B}, in which the authors
found $Z_{\rm Sun}$= $13\pm7$ pc according to the MCs in the range 
of heliocentric distances 0.7--2 kpc. It is probably due to the 
uncertainty of the MC samples located nearby the Sun in their study.  
On the other hand, our independent estimate is in good agreement 
with the recent result of \cite{2016AstL...42....1B}, in which 
they found that the mean value of $Z_{\rm Sun}$ is $16\pm2$ pc 
based on samples of various objects.

In the paper, we only use the limited survey data ($\sim$ 110 
square degrees) to fit the Outer Arm in CO emission. 
The MWISP project will cover the remaining region of the inner
Galaxy in the four or five coming years. We hope that the accumulated
data of the new CO survey will be more helpful in the studying 
of the Outer Arm's structure 
in the first Galactic quadrant.

\subsection{Correlations of the Distant MCs}
In Section 3.3, we show that the distant MCs in the velocity
range of $-1.6\times l$+$13.2\km\ps$ to $-10\km\ps$ trace the Outer Arm
in the Galactic longitude $l=35^{\circ}$ to $l=45^{\circ}$.
In this section, we investigate correlations of the 
distant MCs between the observed parameters, which 
are directly obtained from the CLUMPFIND algorithm (Table 1)
to avoid the uncertainty from other factors
(e.g., the kinematic 
distance from the rotation curve and the estimated mass 
from the CO-to-H$_2$ mass conversion factor $X_{\rm CO}$).
We only consider the correlation of the physical parameters from
\twCO\ because the sample from \twCO\ ($J$=1--0) emission of the
distant MCs is at least four times larger than that 
from \thCO\ ($J$=1--0). Moreover, the
signal-to-noise ratio of the \twCO\ ($J$=1--0) emission is better
than that of \thCO\ ($J$=1--0).

The thick blue line in Figure 8 indicates the relationship between
the size (2R) and the velocity dispersion ($\sigma_{v}$) of the MCs
(the so-called size-linewidth relation):
$\sigma_{v} (\km\ps) \sim \rm {size}^{0.42} (\rm {arcmin})$.
We find that the correlation is not strong. The scattering of the
velocity dispersion is large ($\sim$0.2$\km\ps$--2$\km\ps$)
in the small size range ($\sim$1--8 arcmin).
The small dynamical range, the limited observation sensitivity,
and the non-thermal motions are probably responsible for the poor
fitting \citep[Section 4.3 in][]{2014AJ....147...46Z}.
Although the scattering is large, we also note that the
power-law index of 0.42 is between other works
(\citealp[e.g., 0.33 from][]{1981MNRAS.194..809L}; and
\citealp[0.5 from][]{1987ApJ...319..730S}).
On the other hand, it probably indicates that the velocity dispersion
of MCs is not a simple power-law function of its size
\citep{2001ApJ...551..852H,2009ApJ...699.1092H}.

The left panel of Figure 9 shows a tight correlation
between the CO luminosity ($L_{\rm CO}$) and the virial
mass ($M_{\rm virial}$):
$M_{\rm virial} \sim L_{\rm CO}^{0.87}$.
The power-law index of 0.87 is in good agreement with that of
0.81 from \cite{1987ApJ...319..730S}.
On the other hand, a tight relation between the CO integrated 
intensity ($W_{\rm CO}$) and the virial mass is also found:
$M_{\rm virial} \sim W_{\rm CO}^{2.11}$ with a good correlation 
coefficient of 0.98. The good correlation between the virial mass
and the column density (right panel in Figure 9) is probably related to 
the dependence of $\sigma_{v}/R^{0.5}$ on the column density of MCs 
\citep[see Figures 7 and 8 in][]{2009ApJ...699.1092H}.

The thick blue line in Figure 10 displays the power-law relation
between the size ($2R$) and the CO luminosity ($L_{\rm CO}$)
of the MCs: $L_{\rm CO}\sim \rm {size}^{2.74}$.
We find that the fit of this power-law relation is 
excellent with a correlation coefficient of 0.92.
If we use the weighted CO luminosity to fit it,
a power-law relation of $L_{\rm CO}\sim \rm {size}^{2.41}$
also describes the data.
We emphasize that the above power-law index 2.41 is consistent
well with the result from the optically thin GRS \thCO\ ($J$=1--0) data
(\citealp[a power-law index of 2.36 in ][]{2010ApJ...723..492R},
also see Figure 1 in their paper).
It probably indicates that the distant MCs are also the unresolved
parts of a pervasive fractal structure, which is similar to
the result of \cite{1996ApJ...471..816E}.
The result also shows that the mass of the $large$  MCs traced 
by the \twCO\ ($J$=1--0) luminosity is similar to that traced by
the optically thin \thCO\ ($J$=1--0) emission 
as analyzed by \cite{2010ApJ...723..492R}.
Furthermore, the CO luminosity from the best fit is overestimated for the 
true $L_{\rm CO}$ of MCs with smaller size or lower luminosity.


\section{SUMMARY}
The MWISP project is a new large-scale survey of molecular
gas in \twCO\ ($J$=1--0), \thCO\ ($J$=1--0), and C$^{18}$O
($J$=1--0) emission.
Comparing the data with those of other CO surveys
(\citealp[e.g., GRS \thCO\ ($J$=1--0) survey in][]{2006ApJS..163..145J}
and \citealp[COHRS \twCO\ ($J$=3--2) survey in][]{2013ApJS..209....8D}),
our new CO survey has larger spatial and velocity coverage. In the 
paper, we have presented the result of 110 square degree CO emission 
between $l=$ 34\fdg75 to $l=$ 45\fdg25 and $b=-$5\fdg25 to $b= $5\fdg25
in the velocity interval of 
$-1.6\times l$+$13.2\km\ps$ to $-10\km\ps$
to study the distant MCs of the Milky Way. Based on the new 
unbiased CO survey, the main results are summarized as follows.

1. We have identified over five hundred distant MCs 
according to the \twCO\ ($J$=1--0) emission in the velocity 
range in the 110 square degree 
region. 131 MCs in the \thCO\ ($J$=1--0) emission are also 
identified among these \twCO\ ($J$=1--0) MCs.
The parameters (e.g., position, LSR velocity, peak temperature, 
size, and luminosity; see Tables 1 and 2) of the distant MCs are
presented for the first time.

2, The distribution of the distant MCs in the velocity range is not 
uniform with the Galactic longitude. Four distribution peaks 
of the MCs can be seen at $l=$35\fdg25, 39$^{\circ}$, 
42$^{\circ}$, and 44$^{\circ}$, respectively. 
The interesting implication is that the massive star formation regions
seem to be concentrated in the last three peaks.
The associations between the MCs and the massive star
formation regions can be seen in Table 3.
It also shows that the star formation activity is 
very common in the distant MCs.

3. The distant MCs seem to be concentrated within a narrow
Galactic latitude range along the Galactic longitude.
Most of them are within the region of $b=-$0\fdg5 and $b=$1\fdg5.
We find that the distribution of the distant MCs can be 
described by a Gaussian model with Galactic latitude.
Based on the PV map of the MCs along the Galactic longitude, we thus 
suggest that the CO emission of the MCs is from the Outer Arm 
in the first Galactic quadrant. 

4. According to the unbiased CO survey and the Galactic 
rotation curve of \cite{2014ApJ...783..130R}, the heliocentric distance,
the Galactocentric radii, the scale height, and the lower limit to 
the total mass of the Outer Arm in the segment
are about 15.1--15.6 kpc, 9.5--11.4 kpc, 0\fdg6 (or 160 pc at a 
heliocentric distance of 15 kpc), 
and 2.2$\times10^{6}\Msun$, respectively. 
Assuming a constant value of $X_{\rm CO}$=
$2\E{20}$~cm$^{-2}$K$^{-1}$km$^{-1}$s and a heliocentric distance of 15 kpc, 
the mass surface density of
the distant MCs is about $22\Msun$~pc$^{-2}$.

5. We note that the physical mid-plane traced by the distant
CO arm is slightly displaced with respect to the IAU-defined 
plane, $b=0^{\circ}$. 
We find that the Galactic warp plays a role in the region of 
R$_{\rm GC}$ ($\gsim 10$ kpc) and the tilted-plane model is probably
a good approximation in the region near and within the Solar circle.

6. If the mid-plane within the Solar circle is flat, the tilted angle
between the two planes is about 0\fdg072.
And the distance from the Sun to where the
two planes cross is $d_{\rm Crossing}\sim$ 13.69 kpc assuming
$d_{\rm Sgr A^{\star}}=$8.34 kpc.
The location of the Sun, as an upper limit, is thus about 17.1 pc above the
physical mid-plane from the estimate of the tilted angle. 
The offset of the Sun, which is determined independently from the 
view of the large-scale structure of the distant molecular gas, 
is also in agreement with results via the star-counts 
method.

7. The $L_{\rm CO}$--$M_{\rm virial}$ and the $W_{\rm CO}$--$M_{\rm virial}$
relations of the distant MCs,
as well as the size--$L_{\rm CO}$ relation, display
a convincing power-law relationship,
which is consistent with the results of other studies
\citep[e.g.,][]{1987ApJ...319..730S,2010ApJ...723..492R}.

\acknowledgments
The authors acknowledge the staff members of the Qinghai Radio
Observing Station at Delingha for their support of the observations.
We thank the anonymous referee for critical 
comments and suggestions that helped to improve the paper.
The TOPCAT tool \citep{2005ASPC..347...29T} was used while 
preparing the paper.
This work is supported by NSFC grant 11233007.
The work is a part of the Multi-Line Galactic Plane Survey in CO and its
Isotopic Transitions, also called the Milky Way Imaging Scroll
Painting, which is supported by the Strategic Priority Research Program,
the Emergence of Cosmological Structures of the Chinese Academy of 
Sciences, grant No. XDB09000000.

\bibliographystyle{apj}
\bibliography{references}

\begin{figure*}
\includegraphics[trim=5mm 0mm 0mm 200mm,scale=0.95,angle=0]{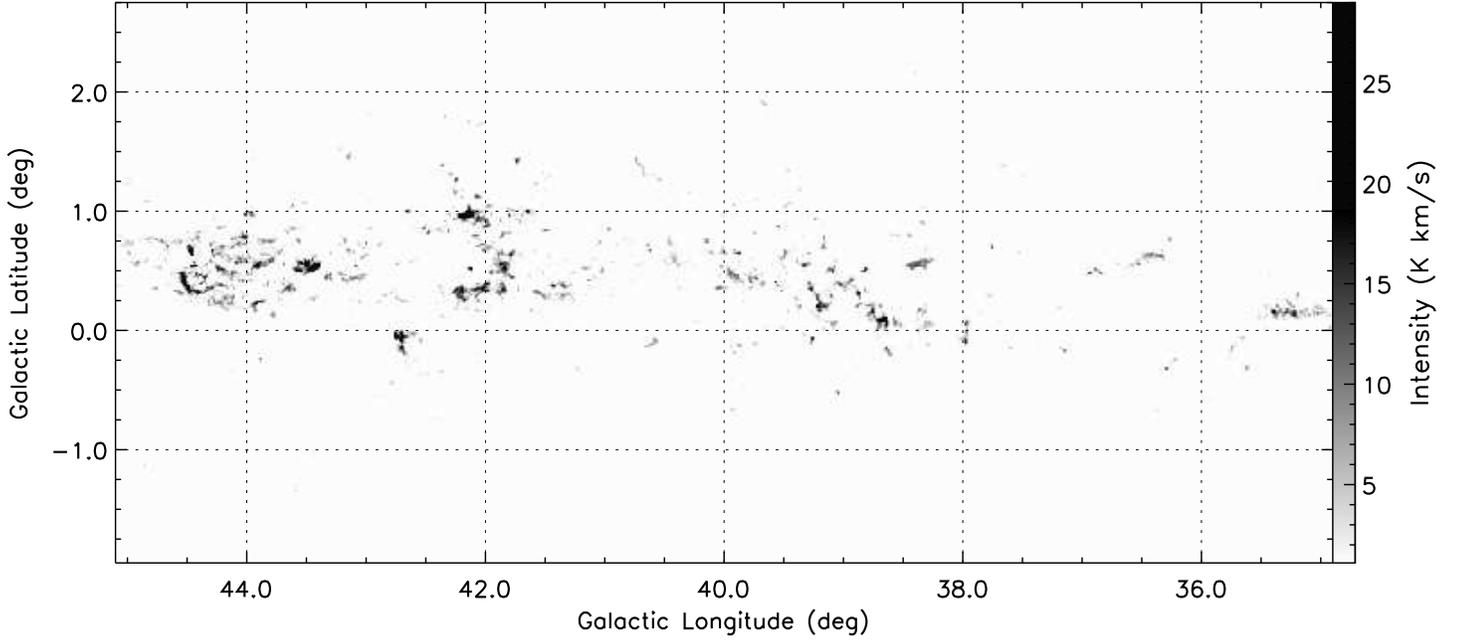}
\caption{
\twCO\ ($J$=1--0) integrated intensity map of 
the MCs on the interval of $-1.6\times l$+$13.2\km\ps$ to 
$-10\km\ps$. Only $>3\times$rms points are accounted for.
\label{fig1}}
\end{figure*}

\begin{figure*}
\includegraphics[trim=5mm 0mm 0mm 200mm,scale=0.95,angle=0]{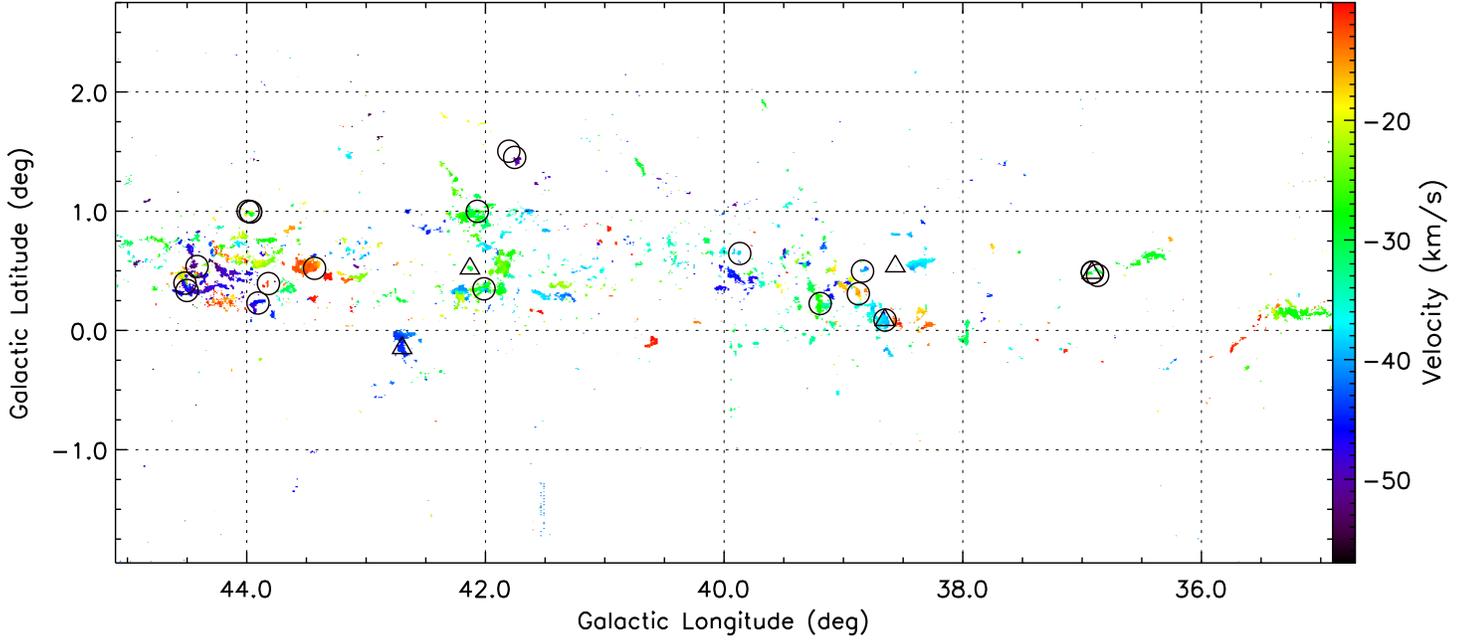}
\caption{
Intensity-weighted mean velocity (first moment) map
of the MCs from \twCO\ ($J$=1--0) emission on the interval of
$-1.6\times l$+$13.2\km\ps$ to $-10\km\ps$. Only $>5\times$rms
points are accounted for.
The black circles and triangles are the positions of the 21 
\HII\ regions and five 6.7 GHz masers from Table 3. The blue stripes at 
$l=$41\fdg5 and $b=-$1\fdg5 are from bad channels at $-41.59\km\ps$ 
(see the text).
\label{fig2}}
\end{figure*}

\begin{figure*}
\includegraphics[trim=0mm 0mm 0mm 0mm,scale=0.3,angle=0]{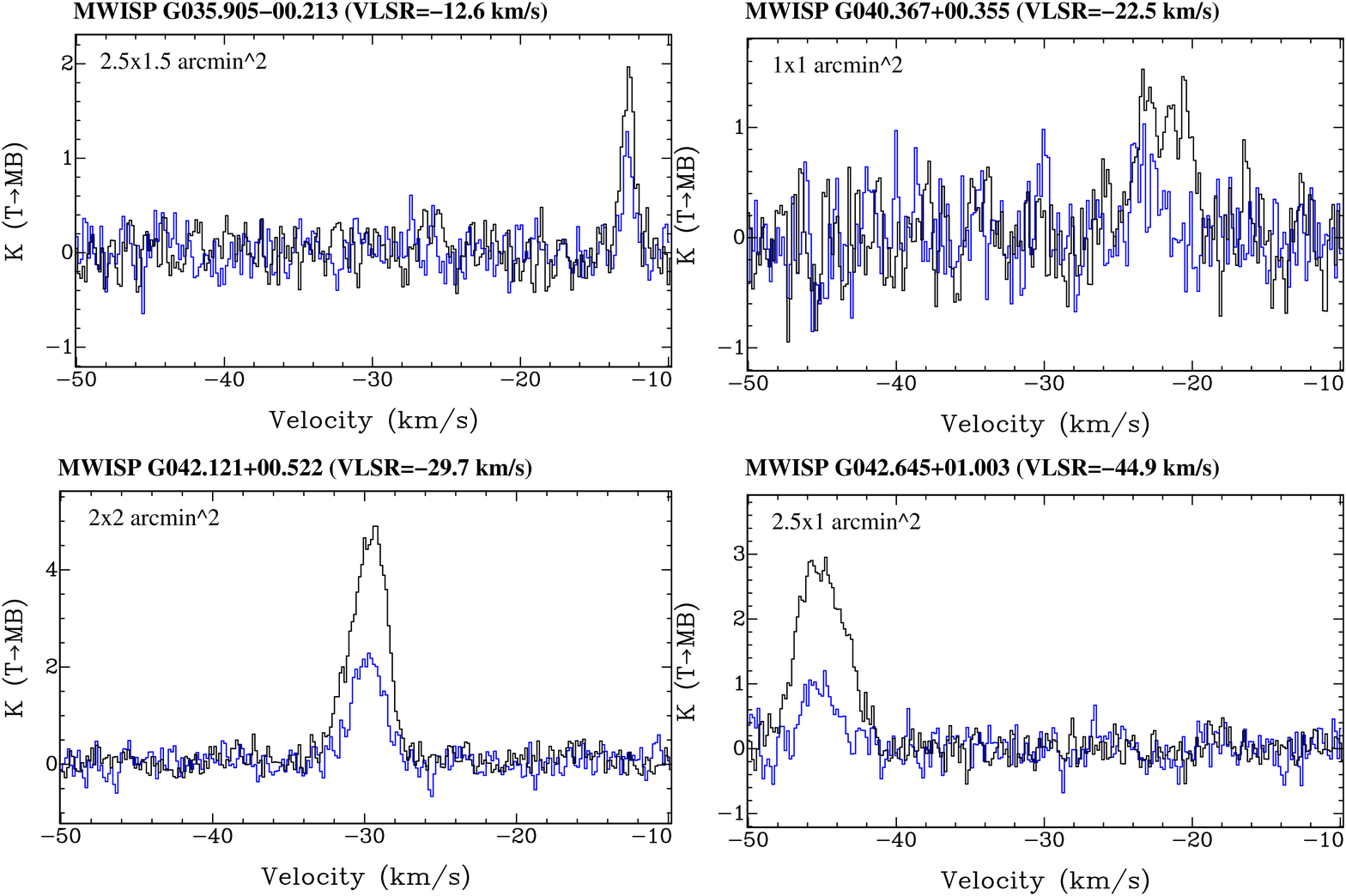}
\caption{Typical \twCO\ ($J$=1--0; black) and \thCO\
($J$=1--0; blue, multiplied by a factor of two)
spectra of the distant MCs. The extraction
area of the spectra is shown in the top-left corner of each panel. 
The $Y$ axis is the main beam temperature of the MCs.
\label{fig3}}
\end{figure*}

\begin{figure*}
\includegraphics[trim=0mm 0mm 0mm 0mm,scale=1.2,angle=0]{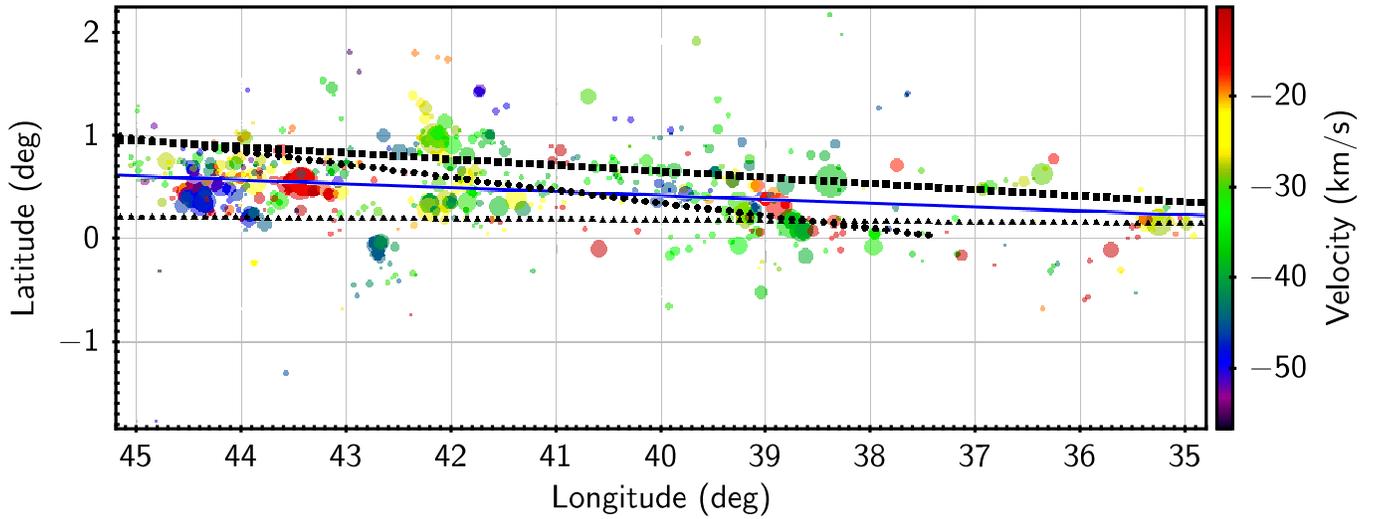}
\caption{
Distribution of the 575 MCs in the Galactic coordinate system.
The filled color circles are 575 MCs in Table~1 and are weighted
with the scale of $L_{\rm CO}^{0.5}$.
The thick blue line (weighted with the luminosity of the MC, 
$b$=0\fdg0377$\times l -$1\fdg0893) shows the best fit by a linear function.
Circle-dotted line: warp model from H\textsc{i} data
\citep[$m=1$ mode, refer to][]{2006ApJ...643..881L};
Box-dotted line: warp model from 2MASS star counts
\citep{2009A&A...495..819R};
Triangle-dotted line: tilted-plane model from 
\cite{1995MNRAS.273..206H}.
\label{fig4}}
\end{figure*}

\begin{figure*}
\centerline{
\includegraphics[trim=5mm 0mm 0mm 0mm,scale=0.6,angle=0]{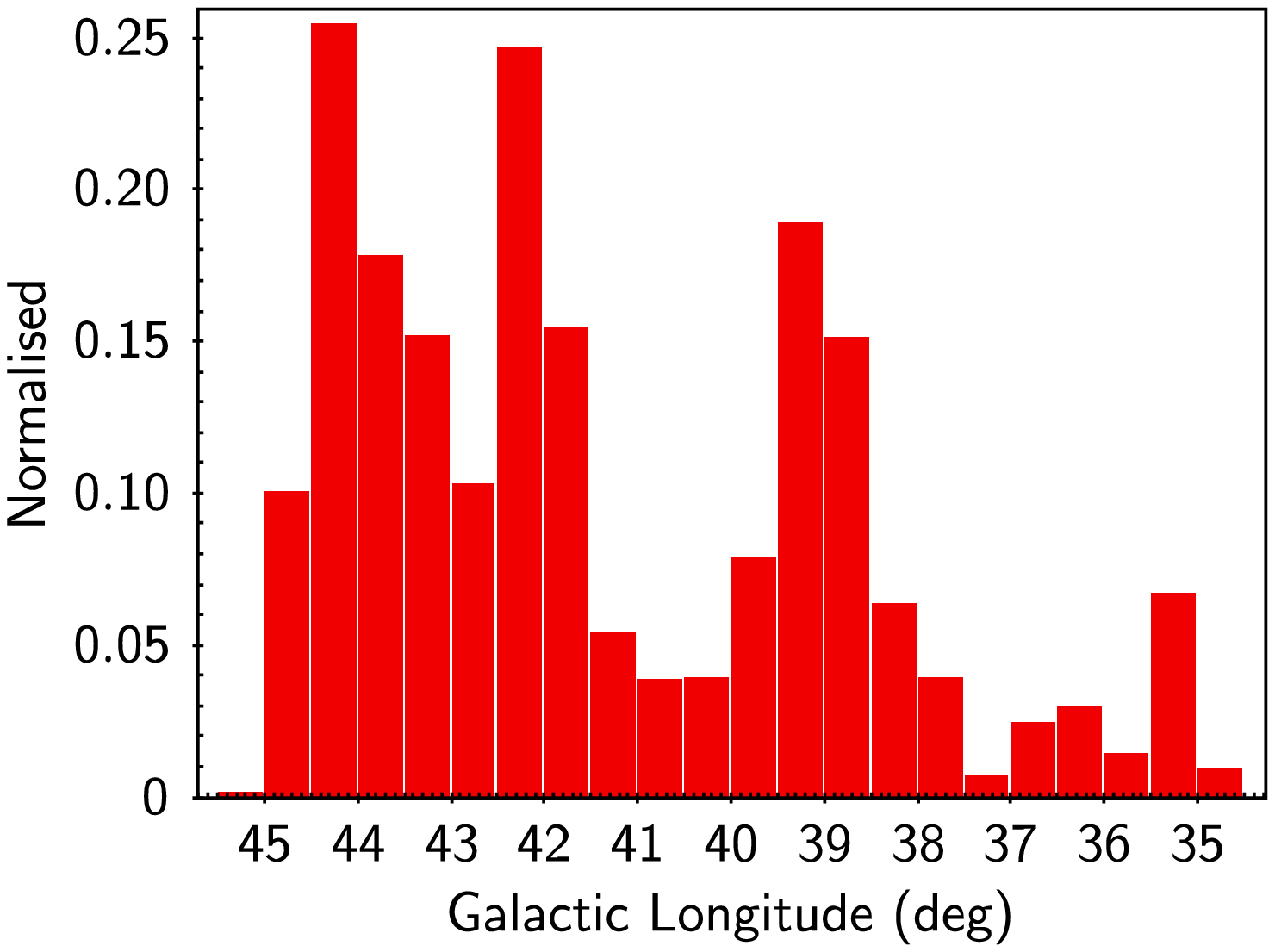}
\includegraphics[trim=5mm 0mm 0mm 0mm,scale=0.6,angle=0]{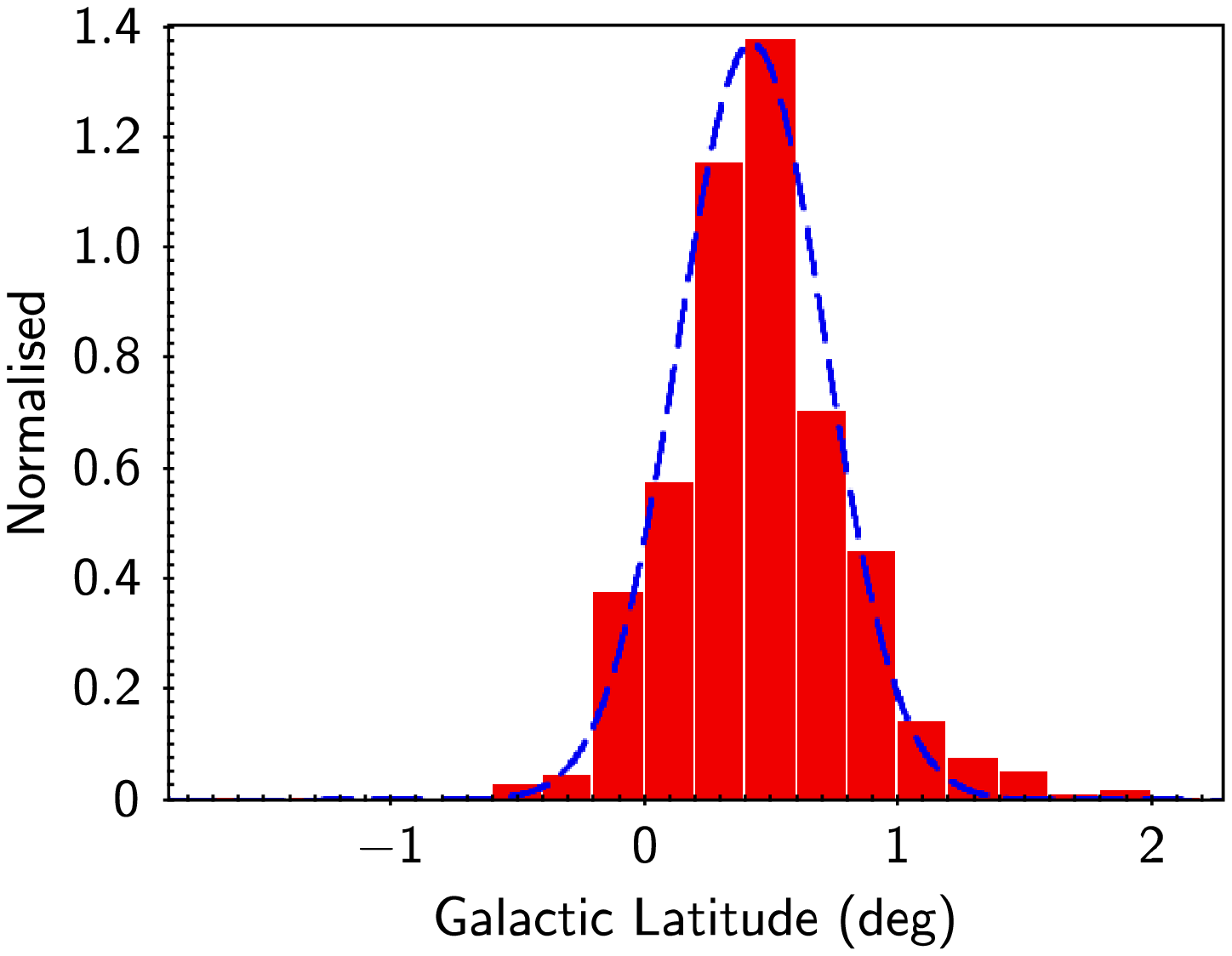}}
\caption{
Histograms of the 575 MCs with the Galactic longitude (the left panel) 
and latitude (the right panel). The dashed blue line in the right 
panel shows the best fit by a Gaussian function with the FWHM of 
$\sim$0\fdg7. 
The y axis is the fraction of the MCs weighted with their 
integrated intensity.
\label{fig5}}
\end{figure*}

\begin{figure*}
\includegraphics[trim=10mm 0mm 0mm 0mm,scale=1.0,angle=0]{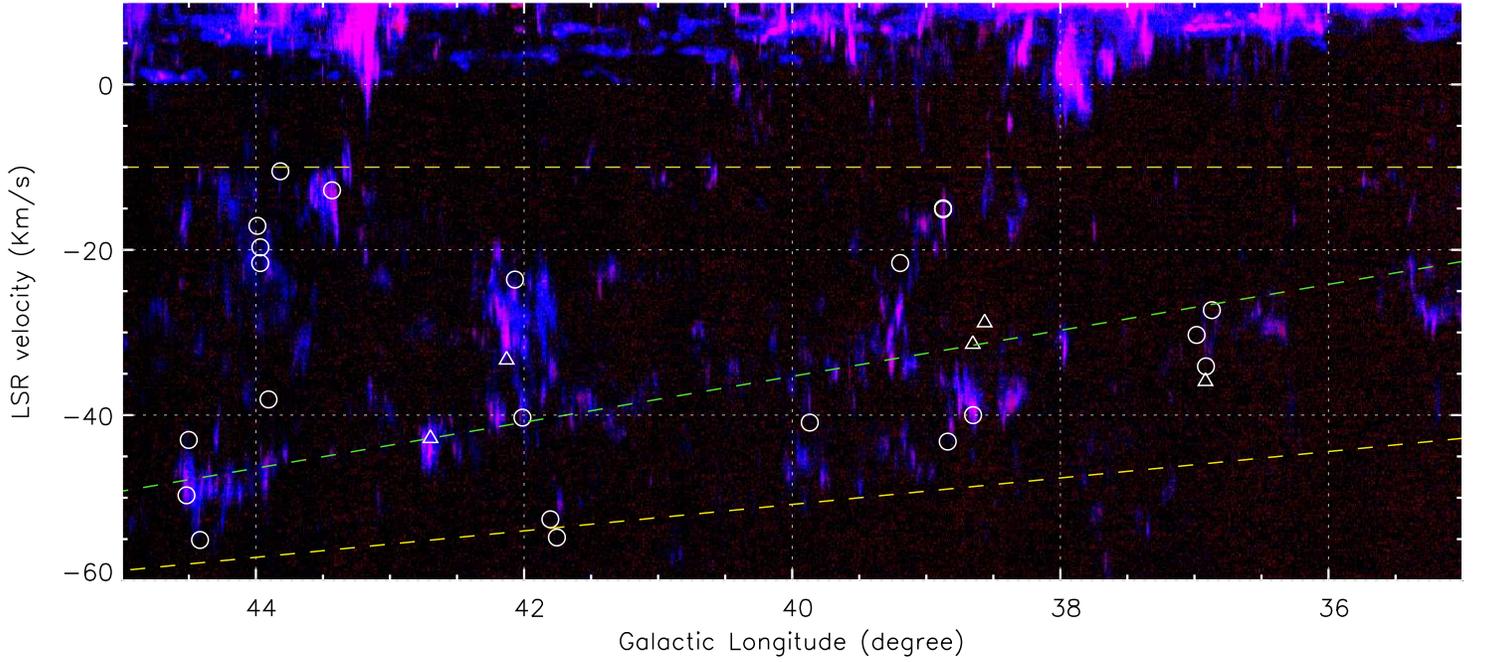}
\caption{
PV diagram of the distant MCs with the Galactic longitude in
the Galactic latitude range of $-$0\fdg5 to 1\fdg5.
The blue and red emission is from the \twCO\ ($J$=1--0)
and \thCO\ ($J$=1--0), respectively.
The white circles and triangles are the same as those 
in Figure 2.
The yellow dashed lines present the LSR velocity range of the Outer Arm:
$V_{\rm LSR}=-1.6\times l+13.2\km\ps$ to $-10\km\ps$.
The green dashed line indicates the longitude$-$velocity relation from the 
best fit of the 
\thCO\ ($J$=1--0) emission (excluding MCs near $l\sim44^{\circ}$ and 
$V_{\rm LSR}\sim-20\km\ps$): $V_{\rm LSR}=-2.78\times l+75.9\km\ps$.
\label{fig6}}
\end{figure*}

\begin{figure*}
\includegraphics[trim=0mm 20mm 0mm 0mm,scale=0.7,angle=270]{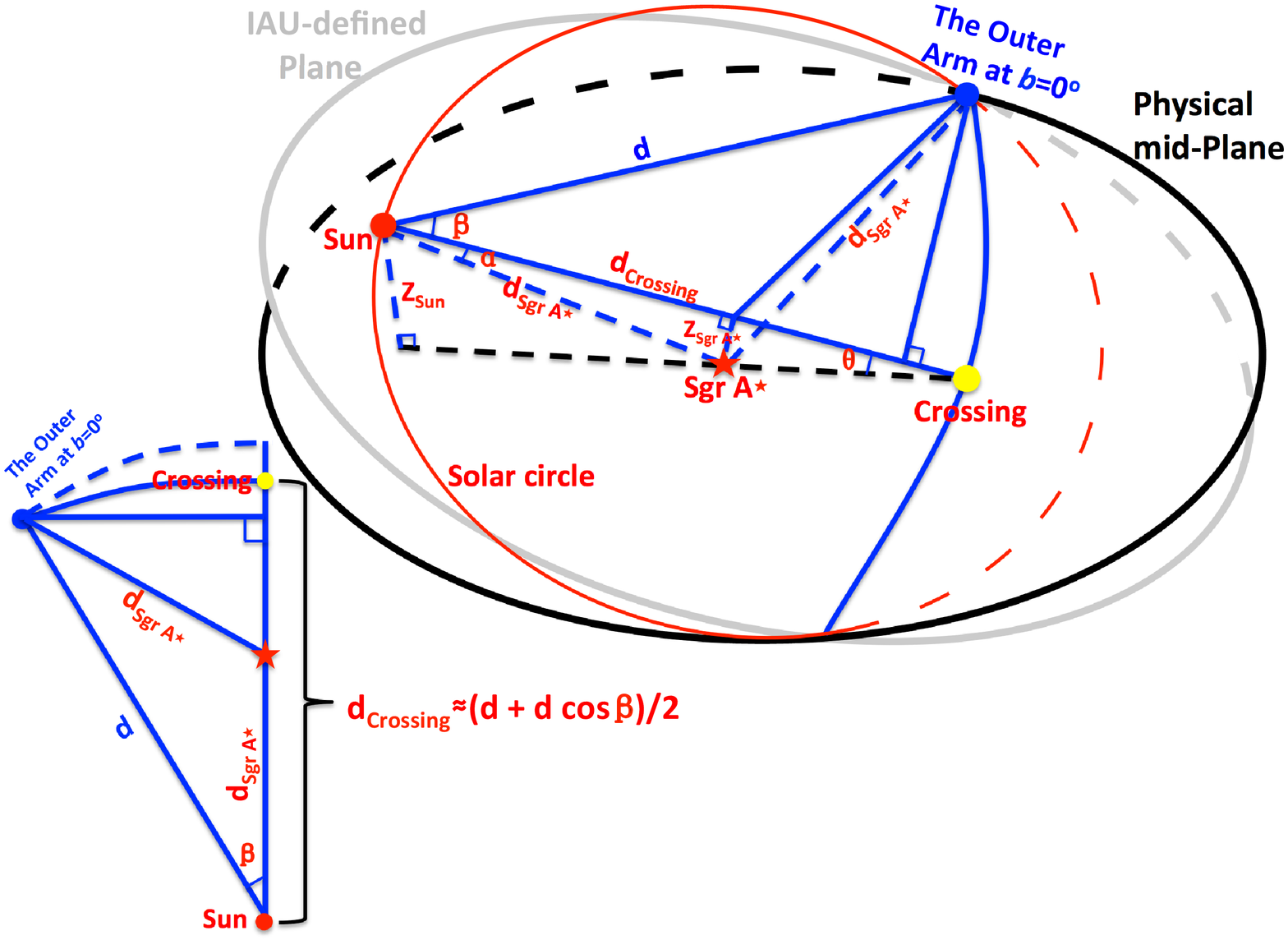}
\caption{
Schematic diagram of the physical mid-plane of the Galaxy with respect
to the IAU-defined plane 
of $b = 0^{\circ}$. Note that the lines and the angles in
the map are not to scale. All solid blue lines are in the 
IAU-defined plane $b = 0^{\circ}$. The red ellipse indicates the 
solar circle. The Outer Arm, which traces the
physical mid-plane in the first Galactic quadrant, passes through
the IAU-defined plane at the direction of $l = \beta$ and
$b = 0^{\circ}$.
$Z_{\rm Sun}$ and $Z_{\rm Sgr A^{\star}}$ represent the height of
the Sun above the physical mid-plane and the offset of the Galactic
center (Sgr A$^{\star}$) below the IAU-defined plane, respectively.
The schematic face-on view of the geometrical relation,
which is the projection onto the IAU-defined plane, is also shown in 
the lower left corner. Thus the value of $d_{\rm Crossing}$
is about 13.69 kpc. 
The $\alpha$, $\beta$, and the tilted angle $\theta$ are $-$0\fdg046,
28\fdg9, and 0\fdg072 (see the text), respectively.
The derived $Z_{\rm Sun}$ is about 17.1 pc.
In the calculation, the distance from the Sun to the Galactic center
is $d_{\rm Sgr A^{\star}}=$ 8.34 kpc.
\label{fig7}}
\end{figure*}

\begin{figure*}
\includegraphics[trim=0mm 0mm 0mm 0mm,scale=1.2,angle=0]{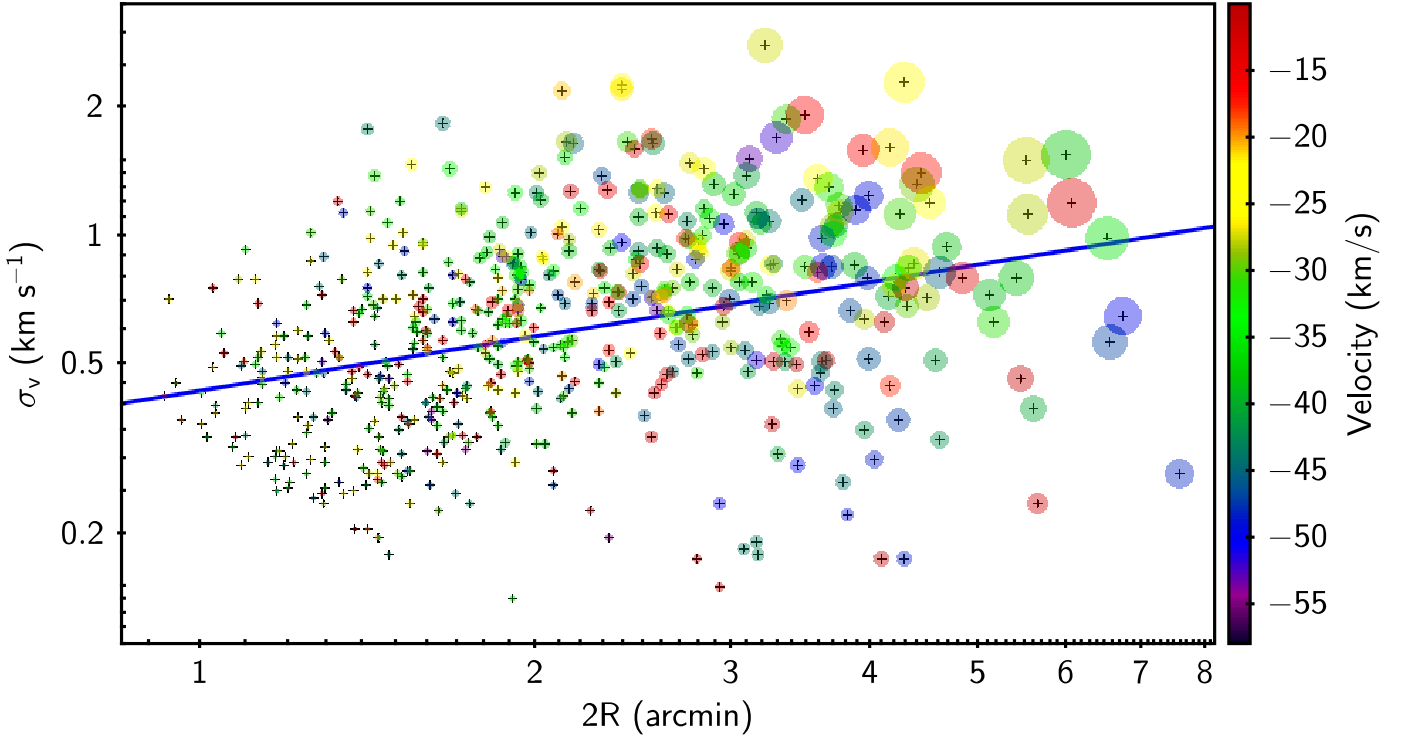}
\caption{
MC velocity dispersion ($\sigma_{v}$=$\Delta V_{\rm FWHM}$/2.355) 
as a function of size ($2R$=2(Area/3.14)$^{1/2}$) for the 575 
clouds in Table~1. The black crosses and the filled circles are 
for the fitted position and the scaled size of the MCs, respectively.
The thick blue line 
($\sigma_{v}\sim \rm {size}^{0.42}$, correlation coefficient (c.c.)=0.34) 
indicates the best fit by a power-law correlation.
\label{fig8}}
\end{figure*}

\begin{figure*}
\centerline{
\includegraphics[trim=5mm 0mm 0mm 0mm,scale=0.65,angle=0]{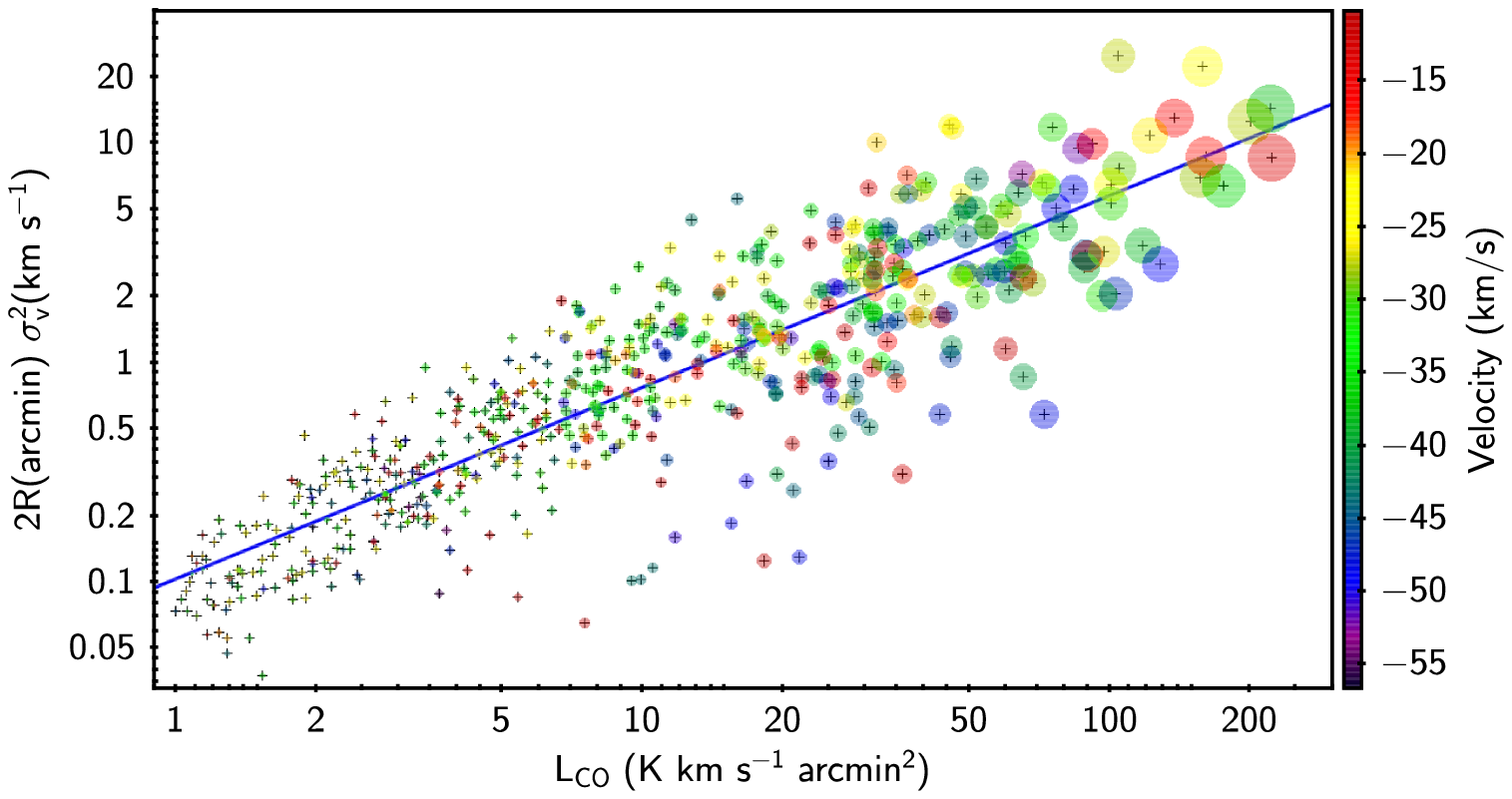}
\includegraphics[trim=-5mm 0mm 0mm 0mm,scale=0.65,angle=0]{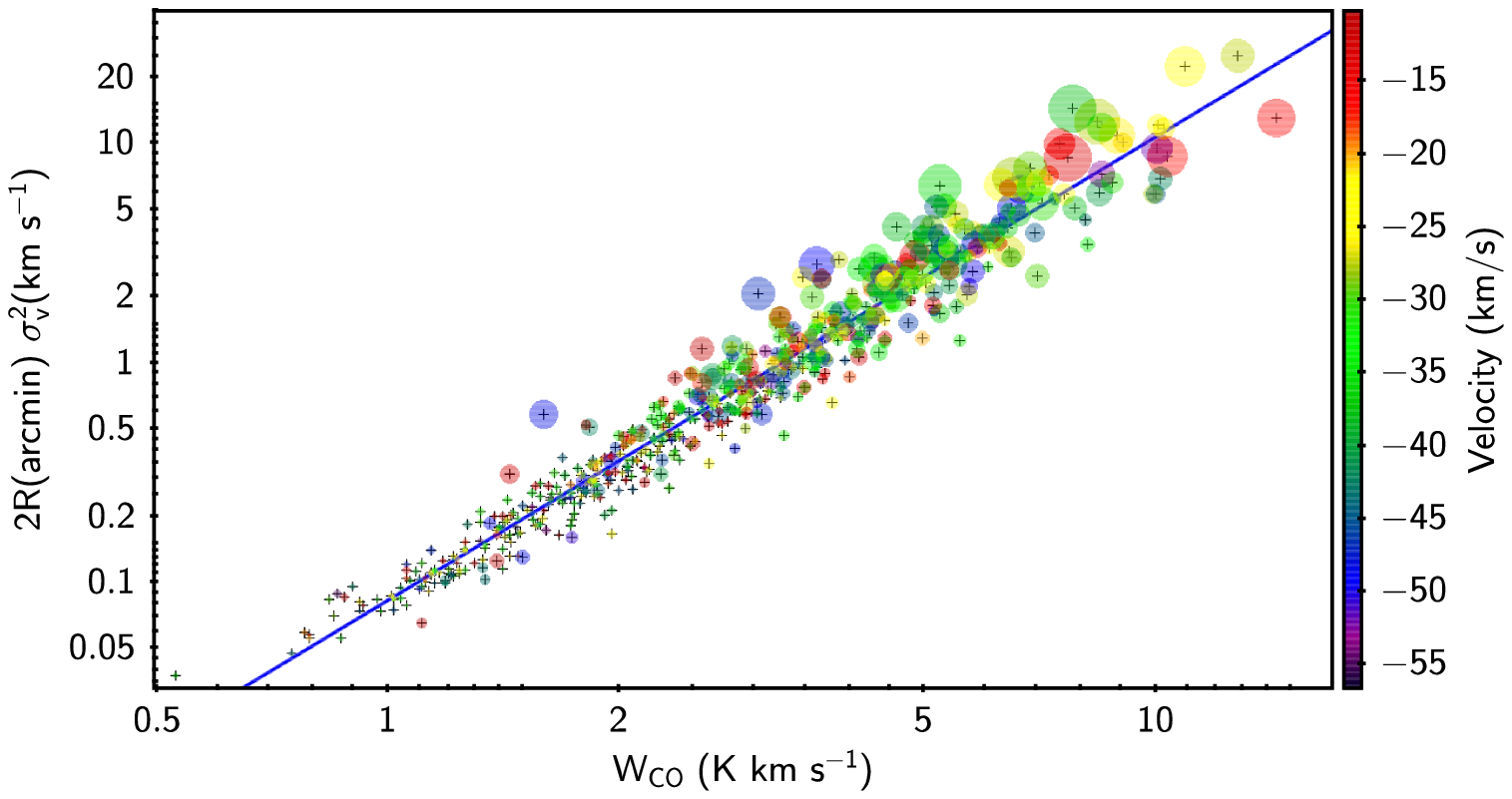}}
\caption{
Left: the CO luminosity ($L_{\rm CO}$)--virial mass 
($M_{\rm virial}\sim 2R \times \sigma_{v}^2$) 
relation for the 575 MCs. The black crosses
and the filled circles are the same as those in Figure 8.
The thick blue line 
($M_{\rm virial}\sim L_{\rm CO}^{0.87}$, c.c.=0.87) 
indicates the best fit by a power-law correlation.
Right: the CO intensity ($W_{\rm CO}$)--virial mass ($M_{\rm virial}$)
relation. The thick blue line 
($M_{\rm virial}\sim W_{\rm CO}^{2.11}$, c.c.=0.98)
indicates the best fit by a power-law correlation.
\label{fig9}}
\end{figure*}

\begin{figure*}
\includegraphics[trim=0mm 0mm 0mm 0mm,scale=1.2,angle=0]{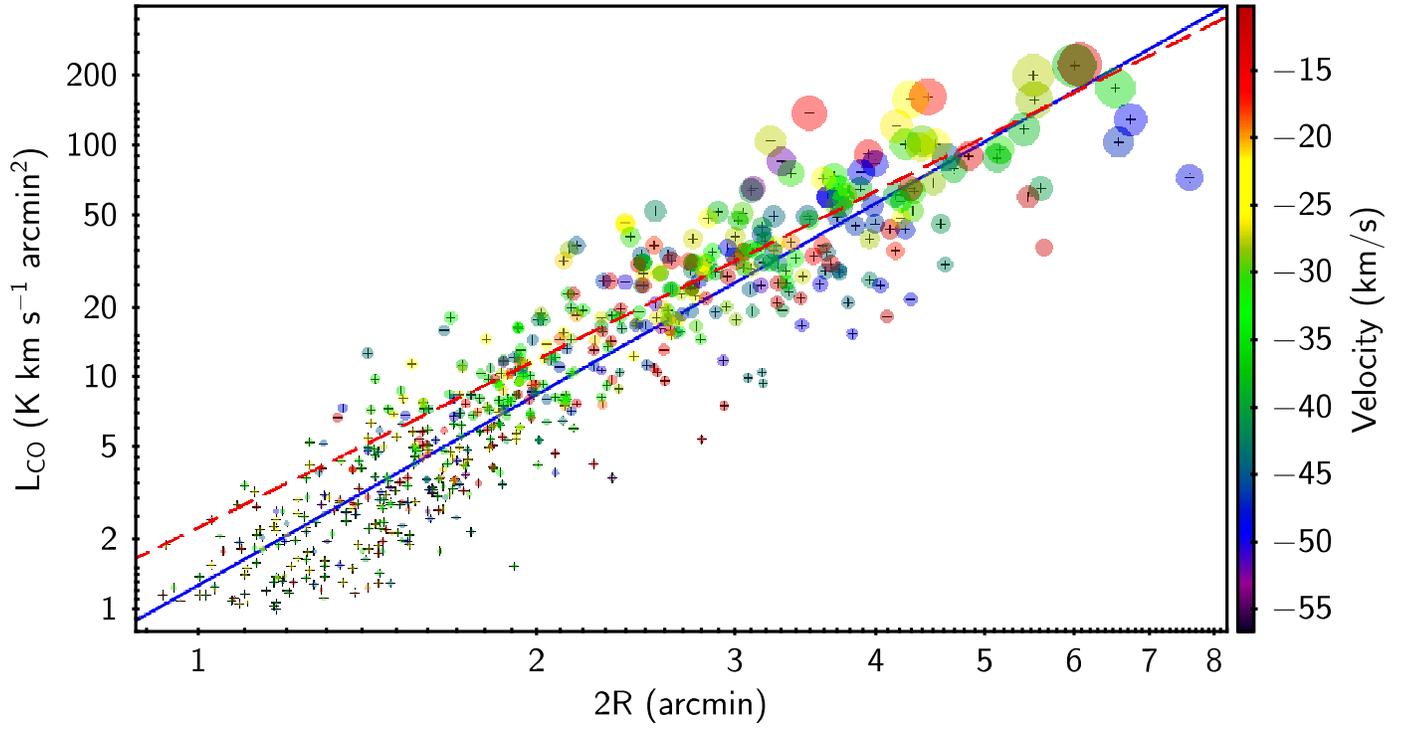}
\caption{
Correlation between the size ($2R$) and the CO luminosity ($L_{\rm CO}$) 
of the 575 MCs. The black crosses
and the filled circles are the same as those in Figure 8.
The thick blue ($L_{\rm CO}\sim \rm {size}^{2.74}$, c.c.=0.92)
and dashed red (weighted with the luminosity of the MC, 
$L_{\rm CO}\sim \rm {size}^{2.41}$, c.c.=0.90) 
lines indicate the best fit by a power-law correlation. 
\label{fig10}}
\end{figure*}

\clearpage
\begin{deluxetable}{ccccccccc}
\tabletypesize{\tiny}
\tablecaption{Parameters of 575 Molecular Clouds Identified 
from \twCO ($J$=1--0) Emission}
\tablehead{
\colhead{\begin{tabular}{c}
ID  \\
    \\
(1)  \\
\end{tabular}} &
\colhead{\begin{tabular}{c}
$l$  	          \\
($^{\circ}$)      \\
(2)  \\
\end{tabular}} &
\colhead{\begin{tabular}{c}
$b$   		  \\
($^{\circ}$)      \\
(3)  \\
\end{tabular}} &
\colhead{\begin{tabular}{c}
$V_{\rm LSR}$      \\
($\km\ps$) 	   \\
(4)  \\
\end{tabular}} &
\colhead{\begin{tabular}{c}
$\Delta V_{\rm FWHM}$     \\
($\km\ps$) 	            \\
(5)  \\
\end{tabular}} &
\colhead{\begin{tabular}{c}
$T_{\rm peak}$   \\
(K) 		 \\
(6)  \\
\end{tabular}} &
\colhead{\begin{tabular}{c}
Area	         \\
(arcmin$^2$)     \\
(7)  \\
\end{tabular}} &
\colhead{\begin{tabular}{c}
$W_{\rm ^{12}CO}$   		     \\
(K $\km\ps$)                 \\
(8)  \\
\end{tabular}} &
\colhead{\begin{tabular}{c}
$L_{\rm ^{12}CO}$   				\\
(K $\km\ps$ arcmin$^2$)                 \\
(9)  \\
\end{tabular}}
}
\startdata
001  &  34.913  &   0.025  &  -23.61  &  2.22  &  2.81  &   1.75  &   3.92  &    6.88 \\
002  &  34.988  &   0.191  &  -24.71  &  1.08  &  2.68  &   1.47  &   1.86  &    2.73 \\
003  &  34.991  &   0.147  &  -27.85  &  1.50  &  5.66  &  12.23  &   3.24  &   39.60 \\
004  &  35.043  &   0.197  &  -12.39  &  0.80  &  2.60  &   2.33  &   1.38  &    3.23 \\
005  &  35.082  &   0.137  &  -26.66  &  1.30  &  2.23  &   1.07  &   2.12  &    2.27 \\
006  &  35.086  &   0.041  &  -18.33  &  0.85  &  2.28  &   1.86  &   1.45  &    2.70 \\
007  &  35.136  &   0.164  &  -26.70  &  2.42  &  3.28  &   4.08  &   4.45  &   18.17 \\
008  &  35.181  &   0.303  &  -38.35  &  2.51  &  2.73  &   2.74  &   4.29  &   11.75 \\
009  &  35.248  &   0.152  &  -28.02  &  2.64  &  5.59  &  24.06  &   6.51  &  156.68 \\
010  &  35.252  &   0.232  &  -22.39  &  2.00  &  4.36  &   8.33  &   4.44  &   36.93 \\
011  &  35.287  &   0.090  &  -25.66  &  0.65  &  2.28  &   1.64  &   1.14  &    1.87 \\
012  &  35.374  &   0.186  &  -14.85  &  2.64  &  4.53  &   5.44  &   5.87  &   31.94 \\
013  &  35.375  &   0.149  &  -22.65  &  3.19  &  5.71  &  10.12  &   7.08  &   71.68 \\
014  &  35.377  &   0.167  &  -27.45  &  0.65  &  2.41  &   1.58  &   1.14  &    1.81 \\
015  &  35.421  &   0.472  &  -22.28  &  1.41  &  2.39  &   1.18  &   2.51  &    2.96 \\
016  &  35.440  &   0.138  &  -24.43  &  0.59  &  2.34  &   1.48  &   1.01  &    1.49 \\
017  &  35.442  &   0.206  &  -23.16  &  2.38  &  3.08  &   1.82  &   4.45  &    8.09 \\
018  &  35.448  &   0.121  &  -12.25  &  1.41  &  2.72  &   2.00  &   2.60  &    5.20 \\
019  &  35.449  &   0.239  &  -24.61  &  0.76  &  2.25  &   1.15  &   1.33  &    1.53 \\
020  &  35.456  &   0.016  &  -11.90  &  1.06  &  2.69  &   2.62  &   1.92  &    5.04 \\
021  &  35.471  &  -0.517  &  -42.42  &  0.68  &  2.35  &   1.11  &   1.19  &    1.32 \\
022  &  35.506  &   0.115  &  -11.03  &  1.00  &  3.28  &   5.14  &   2.03  &   10.44 \\
023  &  35.546  &   0.006  &  -11.09  &  0.48  &  2.63  &   1.49  &   0.79  &    1.17 \\
024  &  35.609  &  -0.326  &  -23.95  &  1.09  &  2.18  &   1.02  &   1.79  &    1.83 \\
025  &  35.613  &  -0.309  &  -26.06  &  2.09  &  3.04  &   3.23  &   3.71  &   12.00 \\
026  &  35.704  &  -0.115  &  -11.69  &  1.47  &  4.58  &  13.29  &   3.25  &   43.26 \\
027  &  35.905  &  -0.213  &  -12.57  &  0.53  &  3.25  &   3.96  &   1.06  &    4.21 \\
028  &  35.927  &  -0.561  &  -11.13  &  1.11  &  3.08  &   1.57  &   1.94  &    3.04 \\
029  &  35.958  &  -0.597  &  -13.27  &  1.77  &  2.26  &   0.83  &   2.92  &    2.42 \\
030  &  36.210  &  -0.242  &  -42.84  &  1.90  &  3.24  &   1.81  &   3.44  &    6.24 \\
031  &  36.250  &  -0.267  &  -37.65  &  1.36  &  2.35  &   1.11  &   2.37  &    2.64 \\
032  &  36.255  &   0.768  &  -16.07  &  2.01  &  6.57  &   4.86  &   5.14  &   24.97 \\
033  &  36.281  &  -0.315  &  -38.17  &  2.72  &  2.91  &   2.32  &   4.88  &   11.29 \\
034  &  36.333  &   0.680  &  -26.65  &  1.52  &  3.26  &   1.91  &   2.69  &    5.13 \\
035  &  36.353  &  -0.680  &  -19.20  &  0.97  &  2.93  &   2.12  &   1.73  &    3.67 \\
036  &  36.368  &   0.631  &  -29.17  &  3.11  &  4.57  &  15.22  &   6.88  &  104.72 \\
037  &  36.430  &   0.169  &  -26.16  &  1.66  &  2.15  &   0.69  &   2.73  &    1.89 \\
038  &  36.441  &   0.578  &  -29.43  &  0.65  &  3.12  &   1.78  &   1.20  &    2.14 \\
039  &  36.534  &   0.562  &  -27.23  &  1.96  &  4.19  &   5.66  &   4.05  &   22.94 \\
040  &  36.618  &   0.529  &  -28.10  &  1.85  &  2.40  &   0.92  &   3.06  &    2.82 \\
041  &  36.639  &   0.464  &  -20.56  &  1.04  &  2.50  &   2.14  &   1.88  &    4.02 \\
042  &  36.703  &   0.567  &  -29.76  &  1.19  &  2.62  &   2.92  &   2.09  &    6.10 \\
043  &  36.716  &  -0.054  &  -44.60  &  1.16  &  2.61  &   0.93  &   2.09  &    1.94 \\
044  &  36.774  &   0.467  &  -26.30  &  0.58  &  1.90  &   1.42  &   0.92  &    1.30 \\
045  &  36.822  &  -0.259  &  -10.48  &  0.68  &  2.69  &   1.69  &   1.30  &    2.20 \\
046  &  36.857  &   0.504  &  -30.51  &  3.90  &  4.88  &   4.61  &   8.78  &   40.46 \\
047  &  36.917  &   0.484  &  -31.68  &  1.90  &  5.57  &   2.91  &   5.57  &   16.20 \\
048  &  36.929  &   0.485  &  -28.19  &  3.90  &  7.38  &   3.58  &   9.94  &   35.62 \\
049  &  37.003  &   0.483  &  -25.55  &  0.92  &  2.51  &   0.95  &   1.49  &    1.41 \\
050  &  37.136  &  -0.165  &  -11.46  &  1.46  &  7.20  &   5.91  &   4.12  &   24.37 \\
051  &  37.174  &  -0.150  &  -37.53  &  1.38  &  3.62  &   2.42  &   2.83  &    6.85 \\
052  &  37.191  &   0.195  &  -33.29  &  2.52  &  2.65  &   1.71  &   4.26  &    7.30 \\
053  &  37.274  &   0.067  &  -14.11  &  0.69  &  2.70  &   1.67  &   1.18  &    1.96 \\
054  &  37.506  &   0.053  &  -31.60  &  0.97  &  2.29  &   1.72  &   1.64  &    2.82 \\
055  &  37.510  &  -0.217  &  -35.93  &  0.83  &  2.51  &   1.58  &   1.40  &    2.22 \\
056  &  37.519  &   0.661  &  -29.68  &  1.33  &  4.22  &   3.66  &   2.90  &   10.63 \\
057  &  37.594  &  -0.147  &  -36.81  &  1.32  &  3.67  &   3.13  &   2.77  &    8.68 \\
058  &  37.643  &   1.408  &  -46.60  &  0.61  &  2.92  &   2.03  &   1.21  &    2.45 \\
059  &  37.656  &   1.391  &  -44.24  &  1.01  &  2.70  &   3.17  &   1.69  &    5.36 \\
060  &  37.747  &   0.709  &  -17.34  &  2.31  &  4.61  &   5.85  &   5.40  &   31.62 
\enddata
\end{deluxetable}

\clearpage
\begin{deluxetable}{ccccccccc}
\tabletypesize{\tiny}
\tablecaption{Parameters of 131 Molecular Clouds Identified from 
\thCO ($J$=1--0) Emission}
\tablehead{
\colhead{\begin{tabular}{c}
ID  \\
    \\
(1)  \\
\end{tabular}} &
\colhead{\begin{tabular}{c}
$l$               \\
($^{\circ}$)      \\
(2)  \\
\end{tabular}} &
\colhead{\begin{tabular}{c}
$b$               \\
($^{\circ}$)      \\
(3)  \\
\end{tabular}} &
\colhead{\begin{tabular}{c}
$V_{\rm LSR}$      \\
($\km\ps$)         \\
(4)  \\
\end{tabular}} &
\colhead{\begin{tabular}{c}
$\Delta V_{\rm FWHM}$     \\
($\km\ps$)                  \\
(5)  \\
\end{tabular}} &
\colhead{\begin{tabular}{c}
$T_{\rm peak}$   \\
(K)              \\
(6)  \\
\end{tabular}} &
\colhead{\begin{tabular}{c}
Area             \\
(arcmin$^2$)     \\
(7)  \\
\end{tabular}} &
\colhead{\begin{tabular}{c}
$W_{\rm ^{13}CO}$                 \\
(K $\km\ps$)                 \\
(8)  \\
\end{tabular}} &
\colhead{\begin{tabular}{c}
$L_{\rm ^{13}CO}$                            \\
(K $\km\ps$ arcmin$^2$)                 \\
(9)  \\
\end{tabular}}
}
\startdata
001  &  34.971  &   0.159  &  -27.50  &  0.72  &  1.20  &  1.45  &  0.58  &   0.84 \\
002  &  35.009  &   0.145  &  -28.25  &  1.08  &  1.99  &  2.12  &  1.02  &   2.17 \\
003  &  35.145  &   0.152  &  -27.60  &  1.13  &  1.13  &  1.14  &  0.93  &   1.06 \\
004  &  35.245  &   0.152  &  -28.33  &  2.48  &  1.87  &  5.39  &  2.21  &  11.93 \\
005  &  35.250  &   0.230  &  -22.10  &  0.70  &  1.53  &  1.89  &  0.65  &   1.23 \\
006  &  35.359  &   0.212  &  -13.66  &  1.21  &  1.29  &  1.13  &  0.97  &   1.09 \\
007  &  35.380  &   0.159  &  -22.85  &  3.11  &  1.65  &  2.16  &  2.81  &   6.09 \\
008  &  35.641  &  -0.059  &  -12.59  &  0.76  &  1.20  &  1.04  &  0.59  &   0.61 \\
009  &  35.737  &  -0.148  &  -11.39  &  1.13  &  1.46  &  2.39  &  1.00  &   2.39 \\
010  &  36.254  &   0.763  &  -16.23  &  0.88  &  1.24  &  1.65  &  0.73  &   1.21 \\
011  &  36.344  &   0.633  &  -29.70  &  1.18  &  1.80  &  4.98  &  1.07  &   5.34 \\
012  &  36.516  &   0.564  &  -27.04  &  0.98  &  1.72  &  1.78  &  0.86  &   1.53 \\
013  &  36.833  &   0.488  &  -28.48  &  0.92  &  1.45  &  1.95  &  0.81  &   1.57 \\
014  &  36.879  &   0.518  &  -31.74  &  1.29  &  1.43  &  1.48  &  1.02  &   1.51 \\
015  &  36.919  &   0.488  &  -30.59  &  2.87  &  1.54  &  1.49  &  2.53  &   3.76 \\
016  &  37.745  &   0.704  &  -17.56  &  1.48  &  2.07  &  3.21  &  1.45  &   4.66 \\
017  &  37.963  &   0.064  &  -32.10  &  1.47  &  2.08  &  2.80  &  1.58  &   4.44 \\
018  &  37.970  &  -0.089  &  -31.74  &  2.25  &  1.90  &  2.28  &  2.13  &   4.86 \\
019  &  38.314  &   0.167  &  -17.13  &  0.71  &  1.15  &  1.23  &  0.58  &   0.71 \\
020  &  38.357  &   0.181  &  -18.14  &  0.83  &  1.16  &  1.76  &  0.63  &   1.11 \\
021  &  38.360  &   0.559  &  -38.12  &  2.03  &  1.68  &  5.32  &  1.76  &   9.35 \\
022  &  38.418  &   0.527  &  -40.39  &  1.46  &  1.81  &  2.25  &  1.31  &   2.96 \\
023  &  38.441  &   0.811  &  -38.79  &  1.04  &  1.06  &  1.63  &  0.78  &   1.27 \\
024  &  38.632  &  -0.150  &  -39.75  &  1.98  &  1.66  &  1.70  &  1.79  &   3.04 \\
025  &  38.655  &   0.083  &  -38.84  &  2.42  &  2.22  &  6.58  &  2.30  &  15.13 \\
026  &  38.753  &   0.169  &  -38.05  &  3.17  &  2.54  &  3.43  &  3.27  &  11.21 \\
027  &  38.771  &   0.100  &  -34.12  &  0.85  &  1.26  &  1.27  &  0.67  &   0.86 \\
028  &  38.809  &   0.520  &  -40.98  &  1.90  &  2.24  &  2.42  &  1.97  &   4.76 \\
029  &  38.869  &   0.312  &  -15.86  &  3.51  &  2.83  &  4.70  &  4.01  &  18.84 \\
030  &  38.879  &   0.210  &  -35.50  &  1.15  &  1.36  &  2.16  &  1.00  &   2.17 \\
031  &  38.928  &   0.436  &  -19.53  &  0.97  &  1.15  &  2.22  &  0.76  &   1.69 \\
032  &  38.953  &   0.404  &  -46.58  &  0.48  &  1.02  &  1.65  &  0.37  &   0.60 \\
033  &  39.013  &   0.387  &  -16.76  &  1.20  &  2.16  &  6.49  &  1.24  &   8.06 \\
034  &  39.058  &   0.077  &  -16.48  &  0.86  &  1.45  &  1.60  &  0.76  &   1.22 \\
035  &  39.088  &   0.064  &  -33.58  &  0.83  &  1.17  &  1.64  &  0.65  &   1.07 \\
036  &  39.094  &   0.293  &  -44.99  &  1.42  &  1.32  &  2.05  &  1.26  &   2.57 \\
037  &  39.148  &   0.199  &  -15.98  &  0.89  &  1.57  &  1.64  &  0.75  &   1.24 \\
038  &  39.159  &   0.212  &  -34.27  &  1.14  &  1.15  &  1.16  &  0.87  &   1.01 \\
039  &  39.198  &   0.208  &  -27.45  &  2.07  &  1.53  &  4.76  &  1.79  &   8.51 \\
040  &  39.216  &   0.304  &  -27.36  &  0.92  &  1.69  &  2.39  &  0.80  &   1.91 \\
041  &  39.256  &  -0.068  &  -33.50  &  1.60  &  2.77  &  6.32  &  1.96  &  12.35 \\
042  &  39.265  &   0.352  &  -30.35  &  1.74  &  2.38  &  3.77  &  1.76  &   6.61 \\
043  &  39.298  &   0.554  &  -43.40  &  1.34  &  1.15  &  1.58  &  1.04  &   1.65 \\
044  &  39.318  &  -0.031  &  -32.41  &  0.70  &  1.12  &  1.12  &  0.54  &   0.61 \\
045  &  39.323  &   0.552  &  -26.92  &  1.63  &  1.89  &  2.01  &  1.50  &   3.02 \\
046  &  39.472  &   0.662  &  -32.52  &  1.22  &  1.09  &  1.19  &  0.92  &   1.10 \\
047  &  39.508  &   0.611  &  -33.80  &  0.58  &  1.19  &  1.21  &  0.46  &   0.56 \\
048  &  39.537  &   0.878  &  -38.26  &  0.84  &  1.42  &  1.39  &  0.72  &   0.99 \\
049  &  39.661  &   1.910  &  -28.94  &  0.92  &  1.13  &  2.12  &  0.71  &   1.49 \\
050  &  39.669  &   0.397  &  -34.82  &  1.00  &  1.70  &  3.40  &  0.91  &   3.11 \\
051  &  39.727  &   0.511  &  -33.34  &  0.56  &  1.21  &  1.84  &  0.48  &   0.88 \\
052  &  39.753  &   0.437  &  -47.74  &  1.41  &  1.68  &  2.18  &  1.16  &   2.52 \\
053  &  39.860  &  -0.118  &  -32.89  &  0.96  &  1.27  &  1.00  &  0.79  &   0.79 \\
054  &  39.870  &   0.658  &  -37.95  &  1.89  &  1.52  &  1.54  &  1.65  &   2.55 \\
055  &  39.901  &  -0.175  &  -32.61  &  0.94  &  1.50  &  1.24  &  0.83  &   1.02 \\
056  &  39.972  &   0.658  &  -35.32  &  0.77  &  0.87  &  0.86  &  0.56  &   0.48 \\
057  &  40.017  &   0.540  &  -47.02  &  1.05  &  1.15  &  1.11  &  0.83  &   0.92 \\
058  &  40.044  &   0.357  &  -44.52  &  0.55  &  1.12  &  1.51  &  0.46  &   0.70 \\
059  &  40.135  &   0.753  &  -30.55  &  1.16  &  1.19  &  1.28  &  0.92  &   1.18 \\
060  &  40.586  &  -0.093  &  -11.36  &  1.58  &  2.14  &  3.72  &  1.51  &   5.61
\enddata
\end{deluxetable}

\clearpage
\begin{deluxetable}{ccccccccc}
\tabletypesize{\tiny}
\tablecaption{Molecular Clouds Associated with Massive Star Formation Regions}
\tablehead{
\colhead{\begin{tabular}{c}
Name $^{\mathrm {a}}$  \\
    \\
\end{tabular}} &
\colhead{\begin{tabular}{c}
$l$             \\
($^{\circ}$)      \\
\end{tabular}} &
\colhead{\begin{tabular}{c}
$b$           \\
($^{\circ}$)      \\
\end{tabular}} &
\colhead{\begin{tabular}{c}
$V_{\rm MC}$      \\
($\km\ps$)         \\
\end{tabular}} &
\colhead{\begin{tabular}{c}
Tracers $^{\mathrm {b,c}}$ \\
    \\
\end{tabular}} &
\colhead{\begin{tabular}{c}
$l$               \\
($^{\circ}$)      \\
\end{tabular}} &
\colhead{\begin{tabular}{c}
$b$               \\
($^{\circ}$)      \\
\end{tabular}} &
\colhead{\begin{tabular}{c}
$V_{\rm LSR}$     \\
($\km\ps$)         \\
\end{tabular}} &
\colhead{\begin{tabular}{c}
References $^{\mathrm {d}}$   \\
  	    \\
\end{tabular}}
}
\startdata
MWISP G036.833$+$00.488       & 36.833 &  0.488 & -28.48 & 
\HII\ G036.870$+$00.462  & 36.870 &  0.462 & -27.3 & (4) \\
MWISP G036.879$+$00.518       & 36.879 &  0.518 & -31.74 &
          		 &	  &	   &        & \\
\hline
MWISP G036.919$+$00.488       & 36.919 &  0.488 & -30.59 &
\HII\ G036.914$+$00.489  & 36.984 &  0.489 & -30.3 & (1) \\
			 & 	  &	   &	    &
\HII\ G036.914$+$00.485  & 36.914 &  0.485 & -34.1 & (4) \\
                         &        &        &        &
Maser G36.918$+$0.483  & 36.918 &  0.483 & -35.9 & (6),(9) \\
\hline
MWISP G038.655$+$00.083       & 38.655 &  0.083 & -38.84 &
\HII\ G038.651$+$00.087  & 38.651 &  0.087 & -40.0 & (4) \\
                         &        &        &        & 
Maser G38.653$+$0.088  & 38.653 &  0.088 & -31.4 & (6),(9) \\
\hline
MWISP G038.809$+$00.520       & 38.809 &  0.520 & -40.98 &
\HII\ G038.840$+$00.497  & 38.840 &  0.497 & -43.2 & (4) \\
\hline
MWISP G038.869$+$00.312       & 38.869 &  0.312 & -15.86 &
\HII\ G038.875$+$00.308  & 38.875 &  0.308 & -15.1 & (4) \\
                         &        &        &        &
\HII\ G038.875$+$00.308  & 38.875 &  0.308 & -14.98 & (5) \\
\hline
MWISP G039.198$+$00.208       & 39.198 &  0.208 & -27.45  &
\HII\ G039.195$+$00.226  & 39.195 &  0.226 & -21.6 & (7) \\
\hline
MWISP G039.870$+$00.658       & 39.870 &  0.658 & -37.95 &
\HII\ G039.869$+$00.645  & 39.869 &  0.645 & -40.9 & (4) \\
\hline
MWISP G041.727$+$01.433 & 41.727 &  1.433 & -50.88 &
\HII\ G041.755$+$01.451  & 41.755 &  1.451 & -54.8 & (8) \\
		     &        &        &        &
\HII\ G041.804$+$01.503  & 41.804 &  1.503 & -52.6 & (8) \\
\hline
MWISP G041.982$+$00.342       & 41.982 &  0.342 & -39.95 &
\HII\ G042.012$+$00.349  & 42.012 &  0.349 & -40.3 & (4) \\
\hline
MWISP G042.120$+$00.521       & 42.120 &  0.521 & -29.76 &
Maser G42.13$+$0.52      & 42.13  &  0.52  & -33.3  & (9) \\
\hline
MWISP G042.150$+$00.982       & 42.150 &  0.982 & -29.16 &
\HII\ G042.068$+$00.999  & 42.068 &  0.999 & -23.6 & (8) \\
\hline
MWISP G042.696$-$00.150       & 42.696 & -0.150 & -44.32 &
Maser G42.698$-$0.147    & 42.698 & -0.147 & -42.8 & (6),(9) \\
\hline
MWISP G043.436$+$00.551       & 43.436 &  0.551 & -14.06 &
\HII\ G043.432$+$00.521  & 43.432 &  0.521 & -12.8 & (4) \\
\hline
MWISP G043.909$+$00.239       & 43.909 &  0.239 & -45.74 &
\HII\ G043.906$+$00.230  & 43.906 &  0.230 & -38.1 & (4)\\
\hline
MWISP G043.950$+$00.983       & 43.950 &  0.983 & -29.68 &
\HII\ G043.968$+$00.993  & 43.968 &  0.993 & -21.6 & (4) \\
			 &	  &	   &	    &
\HII\ G043.989$+$01.000  & 43.989 &  1.000 & -17.1 & (8) \\
\hline
MWISP G044.415$+$00.465       & 44.415 &  0.465 & -49.84 &
\HII\ G044.417$+$00.536  & 44.417 &  0.536 & -55.1 & (4) \\
\hline
MWISP G044.480$+$00.343       & 44.480 &  0.343 & -48.99 &
\HII\ G044.501$+$00.335  & 44.501 &  0.335 & -43.0 & (4) \\
MWISP G044.559$+$00.326       & 44.559 &  0.326 & -46.41 &
			 &	  &	   &	    &    \\
\hline  
MWISP G044.518$+$00.396       & 44.518 &  0.396 & -48.15 &
\HII\ G044.518$+$00.397  & 44.518 &  0.397 & -49.7 & (4)\\
\hline
\hline
\hline
MWISP G043.827$+$00.393 $^{\mathrm {e}}$      & 43.827 &  0.393 & -12.17 &
\HII\ G043.818$+$00.393  & 43.818 &  0.393 & -10.5 & (4) \\
\hline
MWISP G043.957$+$00.998 $^{\mathrm {e}}$      & 43.957 &  0.998 & -20.03 &
\HII\ G043.967$+$00.995  & 43.967 &  0.995 & -19.7 $^{\mathrm {f}}$ &  (2),(3) \\
\hline
\enddata
\tablecomments{
$^{\mathrm {a}}$ Named from Galactic coordinates of the MCs from \thCO\ ($J$=1--0)
emission (Table 2). The last two MCs in the table are from  
\twCO\ ($J$=1--0) emission (Table 1). 
$^{\mathrm {b}}$ We use \HII\ regions and 6.7 GHz methanol masers to trace massive 
star formation regions. The position and the velocity of the \HII\ region is 
from the radio continuum emission and the radio recombination line, respectively. 
The position and the velocity of the 6.7 GHz maser is from the 
observations of the maser survey \citep{2011MNRAS.417.2500G,2015MNRAS.450.4109B}.
$^{\mathrm {c}}$ Based on Tables 1 and 2, we do not find the CO counterpart of
the 6.7 GHZ maser G38.565$+$0.538 with the LSR velocity of $-$28.8 $\km\ps$
\citep[Table 1 in][]{2015MNRAS.450.4109B}. However, a point-like MC with
a temperature of $\sim$1.5 K and the LSR velocity of $-$35.40 $\km\ps$ is
detected from the \twCO\ ($J$=1--0) data at the maser's position.
$^{\mathrm {d}}$ (1)~\citealp{1989ApJS...71..469L}; (2)~\citealp{1996AAS..115...81B};
(3)~\citealp{1996ApJS..103..427G}; (4)~\citealp{2011ApJS..194...32A};
(5)~\citealp{2011ApJ...738...27B}; (6)~\citealp{2011MNRAS.417.2500G};
(7)~\citealp{2012ApJ...759...96B}; (8)~\citealp{2015ApJS..221...26A};
(9)~\citealp{2015MNRAS.450.4109B}.
$^{\mathrm {e}}$ The parameters of the MCs are from \twCO\ ($J$=1--0) emission (Table 1).
$^{\mathrm {f}}$ The velocity of the \HII\ region is from the CS ($J$=2--1) observation.
}
\end{deluxetable}

\end{document}